\documentclass[useAMS,usenatbib]{mn2e}

\usepackage {graphicx}
\usepackage {times}

\title[IFS Spectroscopy of E+A Galaxies with the Kyoto3DII]{Integrated field spectroscopy of E+A (post-starburst) galaxies with the Kyoto3DII\footnotemark[0]}
\author[T. Goto et al.]{Tomotsugu Goto$^{1}$
 \thanks{E-mail:tomo@ir.isas.jaxa.jp} , Atsushi Kawai$^{2}$, Atsushi Shimono$^{2}$,Hajime Sugai$^{2}$,\newauthor Masafumi Yagi$^{3}$, and Takashi Hattori$^{4}$ 
\\
$^{1}$ Institute of Space and Astronautical Science,  Japan Aerospace Exploration Agency,
 3-1-1 Yoshinodai, Sagamihara, Kanagawa 229-8510, Japan\\
$^{2}$Department of Astronomy,
Kyoto University, Sakyo-ku, Kyoto 606-8502, Japan\\
$^{3}$National Astronomical Observatory, 2-21-1 Osawa, Mitaka, Tokyo
181-8588,Japan\\
$^{4}$Subaru Telescope,
National Astronomical Observatory of Japan,
650 North A'ohoku Place, Hilo, Hawai'i 96720, U.S.A.
\\
}

\begin{document}
\def\Hg{H$\gamma$}
\def\Hd{H$\delta$}

\date{\today; in original form 2007 October 17}

\pagerange{\pageref{firstpage}--\pageref{lastpage}} \pubyear{2006}

\maketitle

\label{firstpage}

\begin{abstract}

  We have performed a two-dimensional spectroscopy of 
 three nearby E+A (post-starburst) galaxies with the Kyoto3DII integral field spectrograph. In all the cases,
 H$\delta$ absorption is stronger at the centre of the galaxies, but significantly extended in a few kpc scale.  
 For one galaxy (J1656), we found a close companion galaxy at the same
 redshift. The galaxy turned out to be a star-forming galaxy with a strong emission in H$\gamma$.
 For the other two galaxies, we have found that the central post-starburst regions possibly extend toward the direction of the tidal tails.
 Our results are consistent with the merger/interaction origin of E+A galaxies, where the infalling-gas possibly caused by a galaxy-galaxy merging creates a central-starburst, succeeded by a post-starburst (E+A) phase once the gas is depleted.

\end{abstract}

\begin{keywords}
galaxies: evolution, galaxies:interactions, galaxies:starburst, galaxies:peculiar, galaxies:formation
\end{keywords}

\section{Introduction}

 Galaxies with strong Balmer absorption lines without any
 emission in [OII] nor H$\alpha$ are called E+A galaxies.
 The existence of strong Balmer absorption lines shows 
 that E+A galaxies have experienced starburst recently \citep[within a gigayear;][]{2004A&A...427..125G}.  However, these galaxies do not show any sign of on-going star formation as non-detection in the [OII] emission line indicates.  
   Therefore, E+A galaxies have been interpreted as post-starburst galaxies,
 that is,  a galaxy which truncated starburst suddenly \citep{1983ApJ...270....7D,1992ApJS...78....1D,1987MNRAS.229..423C,1988MNRAS.230..249M,1990ApJ...350..585N,1991ApJ...381...33F,1996ApJ...471..694A}. Recent study found that E+A galaxies have $\alpha$-element excess \citep{2007MNRAS.377.1222G}, which also supported the post-starburst interpretation of E+A galaxies. 
 However, the reason why they started starburst, and why they abruptly stopped starburst remain one of the mysteries in the galaxy evolution. 

  At first, E+A galaxies are found in cluster regions, especially at higher redshift \citep{1985MNRAS.212..687S,1986ApJ...304L...5L,1987MNRAS.229..423C,1988MNRAS.235..827B,1991ApJ...381...33F,1995A&A...297...61B,1996MNRAS.279....1B,1998ApJ...498..195F,1998ApJ...507...84M,1998ApJ...497..188C,1999ApJS..122...51D,1999ApJ...518..576P,2003ApJ...599..865T,2004ApJ...617..867D,2004ApJ...609..683T}. Therefore a  cluster specific phenomenon such as the ram-pressure stripping was thought to be responsible for the  violent star formation history of E+A galaxies \citep{1951ApJ...113..413S,1972ApJ...176....1G,1980ApJ...241..928F,1981ApJ...245..805K,1983ApJ...270....7D,1999MNRAS.308..947A,1999PASJ...51L...1F,2000Sci...288.1617Q,2003ApJ...584..190F,2003ApJ...596L..13B,2004PASJ...56..621F}.
 
 However, \citet{2004MNRAS.355..713B} found that low redshift
E+A galaxies are located predominantly in the field environment, suggesting that a physical mechanism that works in the field region is at least partly responsible for these E+A galaxies. 
Recently, \citet{2005MNRAS.357..937G} has shown that E+A galaxies have more close companion galaxies than average galaxies, showing that the dynamical merger/interaction could be the physical origin of field E+A galaxies. Dynamically disturbed morphologies of E+A galaxies also support this scenario \citep{2006astro.ph.12053L,2005MNRAS.359.1557Y}
 To reconcile the situation, independent evidence on the origin of E+A galaxies has been waited.

  Previous work mentioned above has been focused on the investigation of the global/external properties of E+A galaxies, such as the environment of E+A galaxies\citep{2005MNRAS.357..937G}, and the integrated spectra of the E+A galaxies \citep{2004ApJ...617..867D}.
 However, if the physical origin of E+A galaxies is merger/interaction or gas-stripping, these mechanisms should leave traces to the spatial distribution of stellar-populations in E+A galaxies. For example, a centrally-concentrated post-starburst region is  expected to be found in case of the merger/interaction origin\citep[e.g.,][]{1992ARA&A..30..705B}. In contrast, the gas-stripping would create a more uniform, galaxy-wide post-starburst region. Thus, the spatial-distribution of the post-starburst region inside the E+A galaxy contains important and independent clues on the physical origin of E+A galaxies \citep[e.g.,][]{2005MNRAS.359.1421P,2005ApJ...622..260S}.
 In this work, we try to obtain such independent hints on the origin of E+A galaxies by revealing internal structure of E+A galaxies using the Kyoto3DII integrated field spectrograph (IFS).
  We perform spatially-resolved spectroscopy of three nearby E+A galaxies with the Kyoto-3DII IFS. By revealing the spatial distribution of H$\delta$ absorption, we aim to obtain an independent evidence to shed light on the physical origin of E+A galaxies.

  Unless otherwise stated, we adopt the WMAP cosmology: $(h,\Omega_m,\Omega_L) = (0.71,0.27,0.73)$ \citep{2003ApJS..148....1B}.

\section{Sample Selection}\label{sample}

\begin{table*}
 \begin{minipage}{180mm}
  \caption{Target properties}\label{tab:targets}
  \begin{tabular}{@{}clrrcccc@{}}
  \hline
   Object & Redshift & $g_{AB}$ & $r_{AB}$ & Petrosian Radius (arcsec) in $r$  &   exposure time (sec) & Observing date (HST) & $M_r$\\
 \hline
SDSSJ210258.87+103300.6 & 0.093 & 16.14 & 15.37 & 5.15 &   3600$\times$5 & July, 20, 2005 &        -23.00 \\
SDSSJ165648.64+314702.3 & 0.100 & 17.73 &  16.72 & 2.32 &   3600$\times$2+1800 & July, 21, 2005 &  -21.74\\
SDSSJ233712.76-105800.3 & 0.078 & 16.02 &  15.31 & 5.45 &   3600$\times$3 & July, 21, 2005 &       -22.47   \\
\hline
\end{tabular}
\end{minipage}
\end{table*}



 We select our target E+A galaxies from \citet{2005MNRAS.357..937G,goto_DR6}, which presented a catalog of 564 E+A galaxies based on the $\sim$670,000 galaxy spectra of the Sloan Digital Sky Survey \citep{2006ApJS..162...38A}. 

 Briefly, the 564 E+A galaxies satisfy the following criteria: 
\begin{itemize}
 \item H$\delta$ EW $>4$\AA \footnote{Absorption lines have a positive sign throughout this paper.}
 \item {[OII] EW $>-$2.5\AA}
 \item H$\alpha$ EW $>$ $-$3.0\AA
\end{itemize}

  Note that the line measurement was performed on the spectra through the SDSS fiber spectrograph with the diameter of three arcsec. 
  The use of H$\delta$ line is preferred over other hydrogen Balmer lines (e.g., H$\zeta$, H$\epsilon$, H$\gamma$, H$\beta$)
 since the line is isolated from other emission and absorption lines, as well
 as strong continuum features in the galaxy spectrum (e.g.,
 D4000). Furthermore, the lower order Balmer lines (H$\gamma$ and H$\beta$)
 can suffer from significant emission-filling, while the higher order lines (H$\epsilon$ and H$\zeta$) have a low
 signal-to-noise in spectra. 
We stress that our criteria of H$\delta$ equivalent width (EW) $>4$\AA~ is a very strong one compared with previous work.
 Thus, we can select strong post-starburst galaxies, which previous work with a smaller population were not able to find.
 
 Out of the 564 E+A galaxies, we have observed three as described in Section \ref{Oct 12 11:18:49 2007}.

\section{Observation}
\label{Oct 12 11:18:49 2007}

 Observations were carried out on the nights of July 20,21 and 22nd with the Kyoto tridimensional spectrograph II (Kyoto3DII)$-$Integral field spectrograph (IFS) \citep{2004SPIE.5492..651S,2006NewAR..50..358S,2007ApJ...660.1016S} attached to the University of Hawaii 88inch (2.2 meter) telescope (UH88). Unfortunately, no useful data were taken on the 22nd due to the heavy cloud. The Kyoto3DII is a multi-mode spectrograph with four observational modes: Fabry-Perot imager, integral field spectrograph (IFS), filter-imaging modes, and slit spectroscopy. 
When attached to the UH88, the IFS has the field of view of $\sim$14''$\times$16'' with the pixel-scale is 0.43''/pix. We used the No.2 grism and the No.2 filter which cover the wavelength range of  4200-5200\AA~ with the resolution of $R\sim$1200.
 
 Among the 564 E+A galaxies described in Section \ref{sample}, we observed three targets which had a bright magnitude, an appropriate redshift to measure H$\delta$ line, and good visibility on the observing dates. 
 We present the basic properties of the three targets in table \ref{tab:targets}, where measured quantities such as positions, redshift, magnitudes in $g$ and $r$, and Petrosian radius are taken from the SDSS catalog. Here, $R_{90}$ is the radius within which 90\% of the $r$-band Petrosian flux is contained. Magnitudes are de-reddened (for the Galactic extinction) Petrosian AB-magnitude in $g$ and $r$-band. 
 In the panel (a) of Figures \ref{fig:J2102}-\ref{fig:J2337}, we show the $g,r,i$-composite images of the three targets. 
 J1656 has a possible companion galaxy to the east. J2337 has a tidal tail from north-west to south-east. 

The average seeing was $\sim$1.4 arcsec on both nights based on the PSF measured with standard stars. 
 The seeing size corresponds to $\sim$3 lenslets, which are smaller than the spatial scale we probe in Section \ref{Sep 28 14:10:45 2007}.
 Exposure time was 2.5-5 hours per target depending on the target visibility.
Data reduction was carried out using the {\ttfamily mla} IRAF package for Kyoto3DII (Kawai et al. in prep.)
\footnote{http://smoka.nao.ac.jp/about/subaru.jsp},
 which takes care of the shift due to the atmospheric dispersion \citep{1982PASP...94..715F}, and that between the exposures in addition to the basic reduction. We also observed standard stars BD+253941 and BD+284211 for flux calibration. A He-Hg lamp was used for wavelength calibration. 
The Kyoto3DII has a separate fields of view (2.5' away from the target on the UH88) for the simultaneous sky subtraction \citep{2006NewAR..50..358S}.  
During our observation, however, there was slight additional light (suspected sky gradient due to the moon shine) in the sky lenslets. Therefore, we used $\sim$100 lenslets in the targets area (but well away from the target) for the sky subtraction. 

\section{Results}
\label{Sep 28 14:10:45 2007}

\subsection{SDSSJ210258.87+103300.6}

\begin{figure*}
\begin{center}
\includegraphics[scale=0.33]{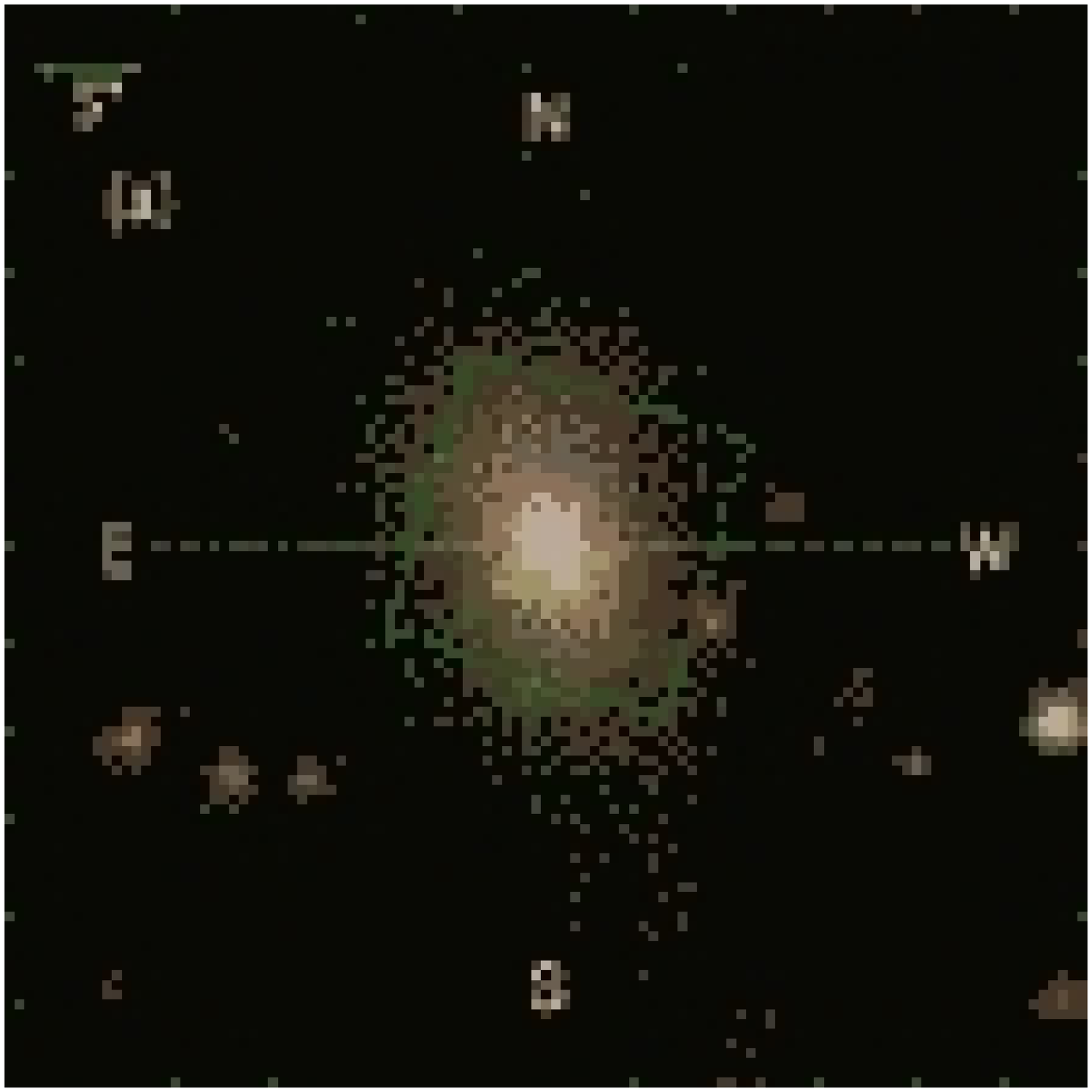}
\includegraphics[scale=0.31]{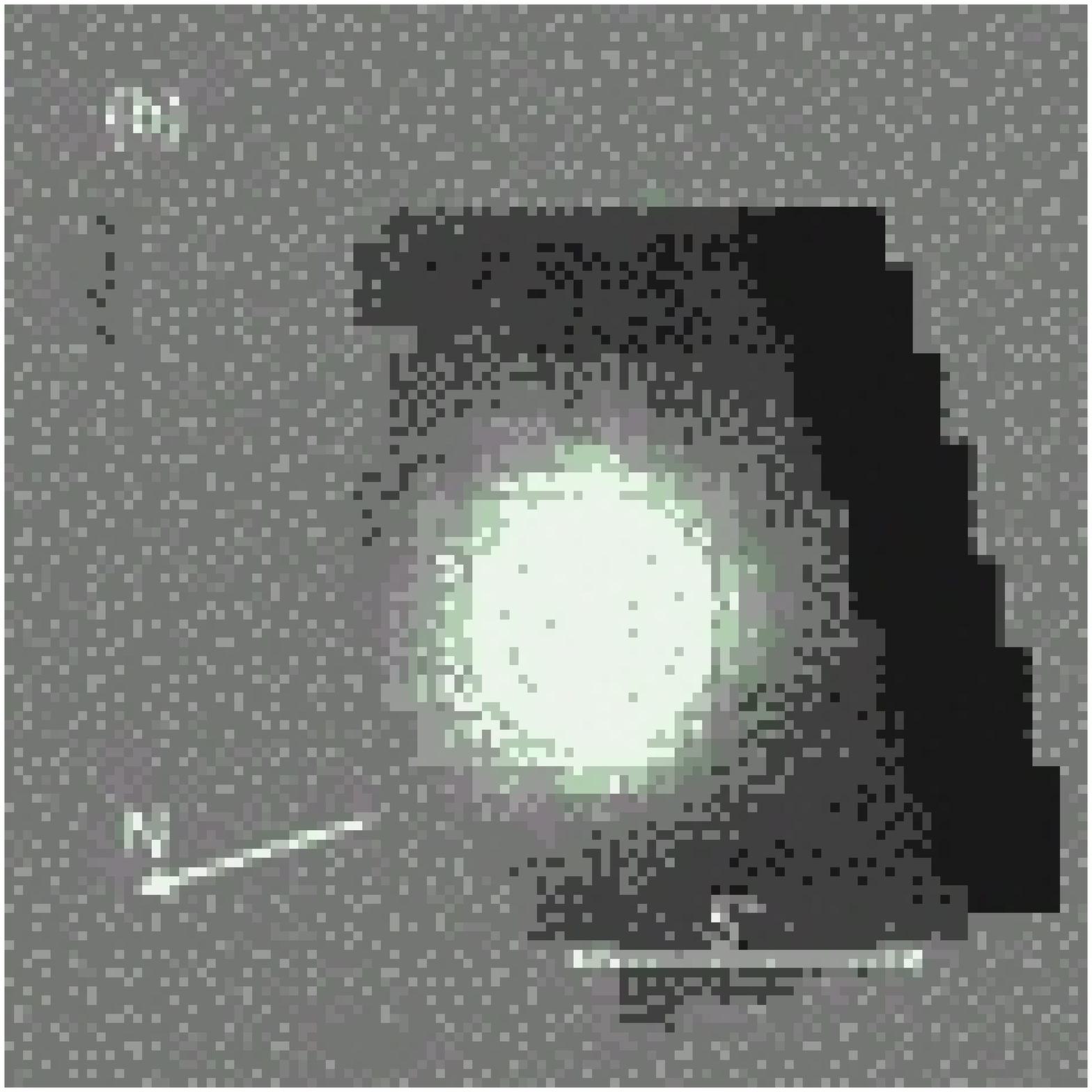}
\includegraphics[scale=0.31]{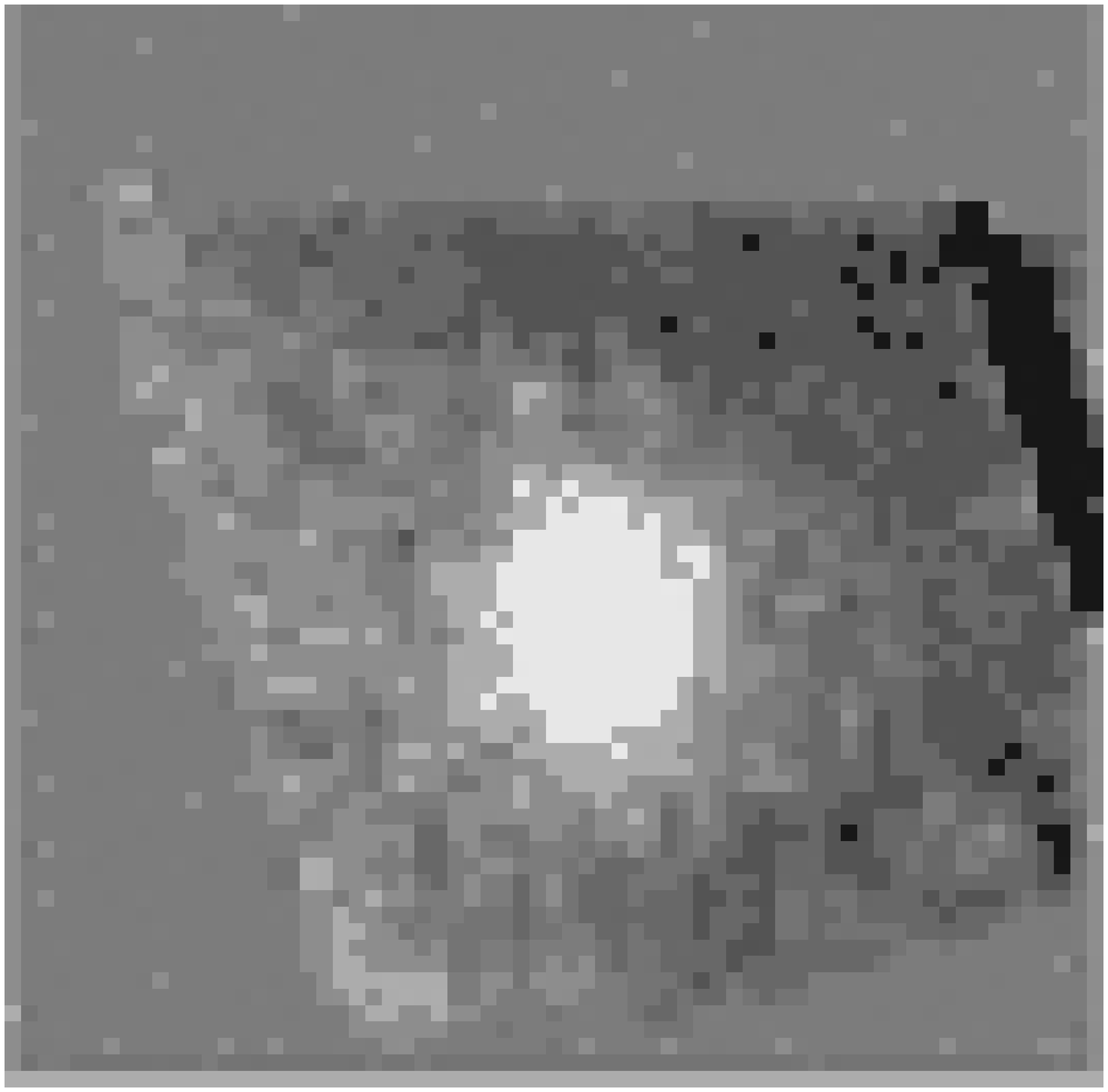}
\includegraphics[scale=0.31]{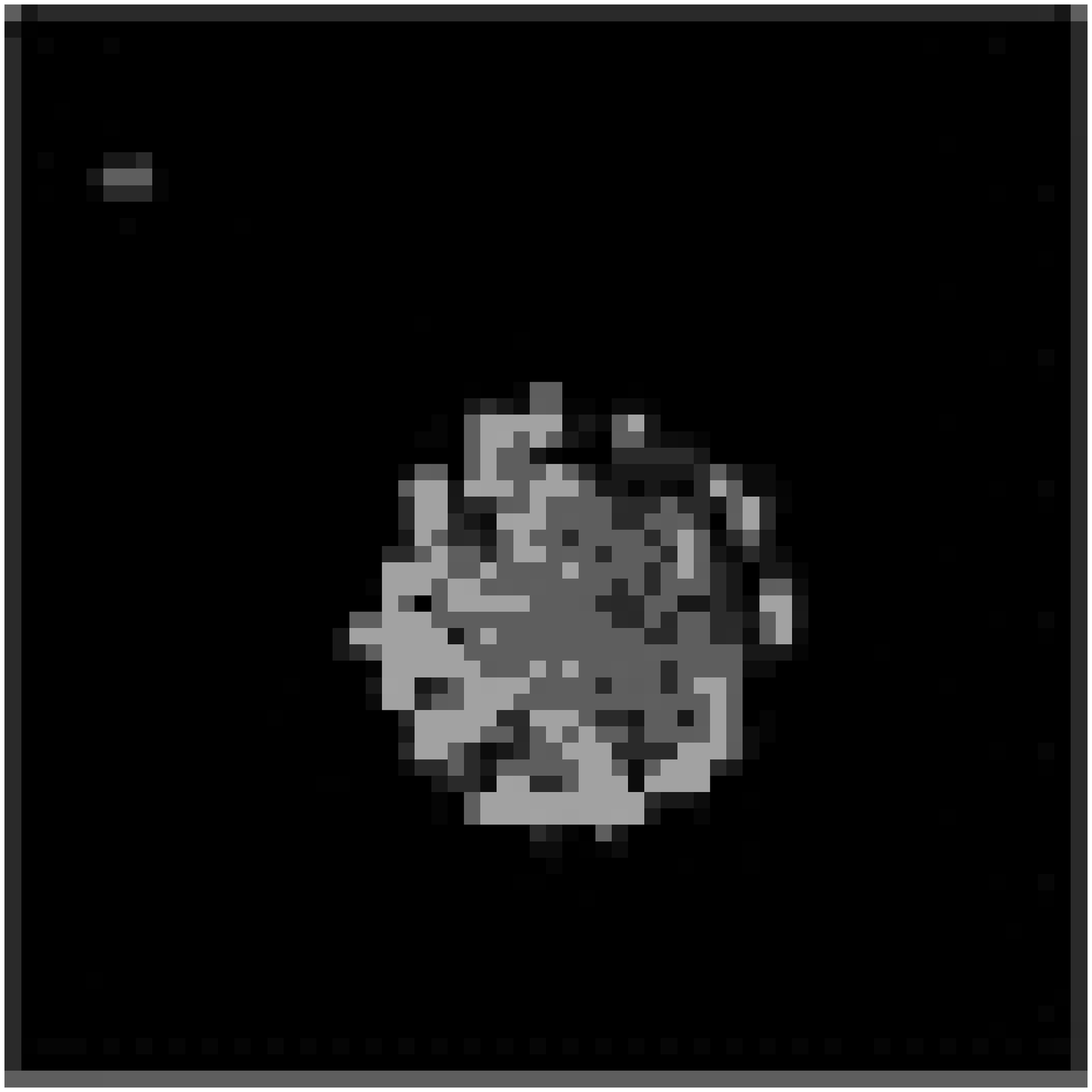}
\includegraphics[scale=0.45]{071002_k3d_radial_plot_kpc.ps_pagesj2102}
\includegraphics[scale=0.45]{071002_k3d_radial_plot_kpc.ps_pagesj2102r}
\end{center}
\caption{Figures for J2102: (a) $g,r,i$-composite image taken from the SDSS website. The kyoto-3DII field-of-view is overlaid.(b) Observed IFS image using the light in all the wavelengths (4200-5200\AA). Overplotted circles are with the radii of 1.5,3,4.5 and 6 lenslet. The orientation of the figure is indicated with the arrow. (c) IFS image using the light in H$\delta$ wavelength. The effective data area is lager than that in the panel (b) because of the smaller wavelength range used. (d) H$\delta$-to-continuum ratio.  The scale and orientation are common among panels (b), (c), and (d). (e)  Spectra of the central 1.5 lenslet radius region, and annuli of 1.5-3, 3-4.5, 4.5-6 lenslet regions of J2102 from top to the bottom.  The spectra are shifted to the rest-frame and smoothed using a 5-pixel box.  (f) H$\delta$ (diamonds) and H$\gamma$ (triangles) EWs are plotted against the distance to the galaxy centre.
}\label{fig:J2102}
\end{figure*}

We show the image of SDSSJ210258.87+103300.6 reconstructed from the whole light through the IFS (4200-5200\AA) in the panel (b) of Figure  \ref{fig:J2102}.  Overplotted circles are with the radii of 1.5,3,4.5 and 6 lenslet. 

 In the panel (c), we show the image constructed using only the light through H$\delta$ wavelength (4086-4110\AA~to be exact). 
To see the difference, we take the ratio of H$\delta$ to all the wavelength in the panel (d). See Appendix \ref{Oct  3 10:22:16 2007} for the details in estimating the ratio when both components have large uncertainty.  
Because H$\delta$ is an absorption line, it appears darker when stronger. 
We checked that this method returned a uniform image (ratio) for a standard star (see Appendix \ref{Oct  3 10:21:43 2007} for details).
 In the panel (d), the centre of the galaxy is slightly darker with a little extension to the right, showing that H$\delta$ absorption may be concentrated around the galaxy centre. Note that the pixel values are meaningless outside of the galaxy, where no signal is present other than the noise.
 It is quite striking that the distribution of the post-starburst phenomena (with strong H$\delta$ absorption) is mapped out in two-dimension. The figure demonstrates the ability of the IFS observation to disentangle spacial and spectral information. 
 
 The trend becomes clearer in the spectrum. 
In the panel (e) of Figure \ref{fig:J2102}, we show the spectra of the central 1.5 lenslet radius region, and annuli of 1.5-3, 3-4.5, 4.5-6 lenslets regions  of J2102 from top to the bottom. The spectra are shifted to the rest-frame and smoothed using a 5-pixel box.  The spectrum from the outermost annulus (4.5-6 lenslets of radius) is approaching to zero and not reliable.
 In the panel (e), it is noted that H$\delta$ and H$\gamma$ absorptions become deeper towards the centre of the galaxy.
 To quantify this, we have measured H$\delta$ and H$\gamma$ EW in these spectra using the flux summing method described in \citet{2003PASJ...55..771G}. Results are in the panel (f) of Figure \ref{fig:J2102}, where  H$\delta$ EWs are plotted with diamonds and   H$\gamma$ EWs are with triangles. H$\delta$ EW is the strongest at the centre, but remains strong as far as $\sim$3 kpc away. Note that the seeing size is $\sim$3 lenslets, and smaller than the extension of the strong H$\delta$ absorption.  H$\gamma$ EWs are not so strong as  H$\delta$, but show a similar trend.

%

%

\subsection{SDSSJ165648.64+314702.3}
\begin{figure*}
\begin{center}
\includegraphics[scale=0.33]{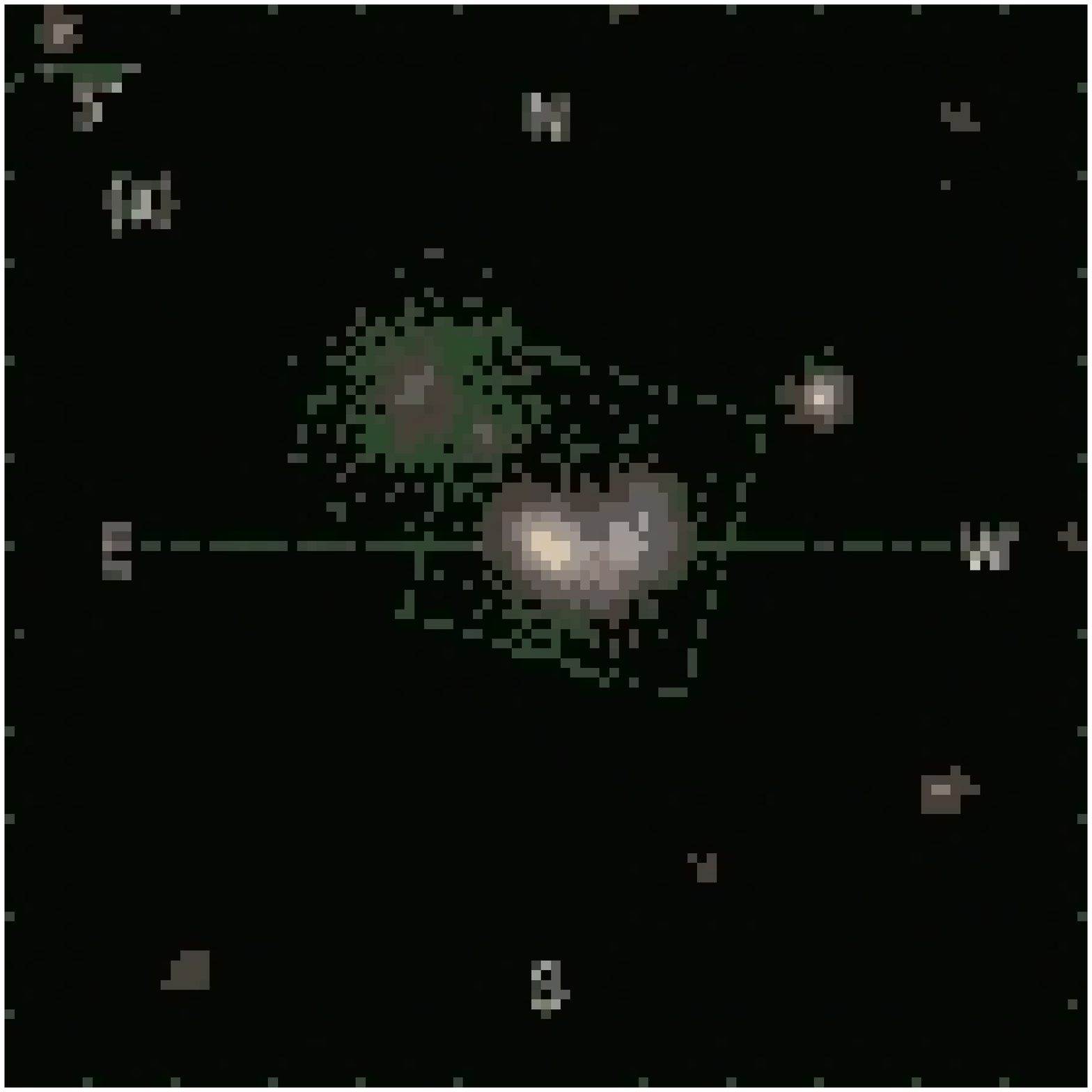}
\includegraphics[scale=0.31]{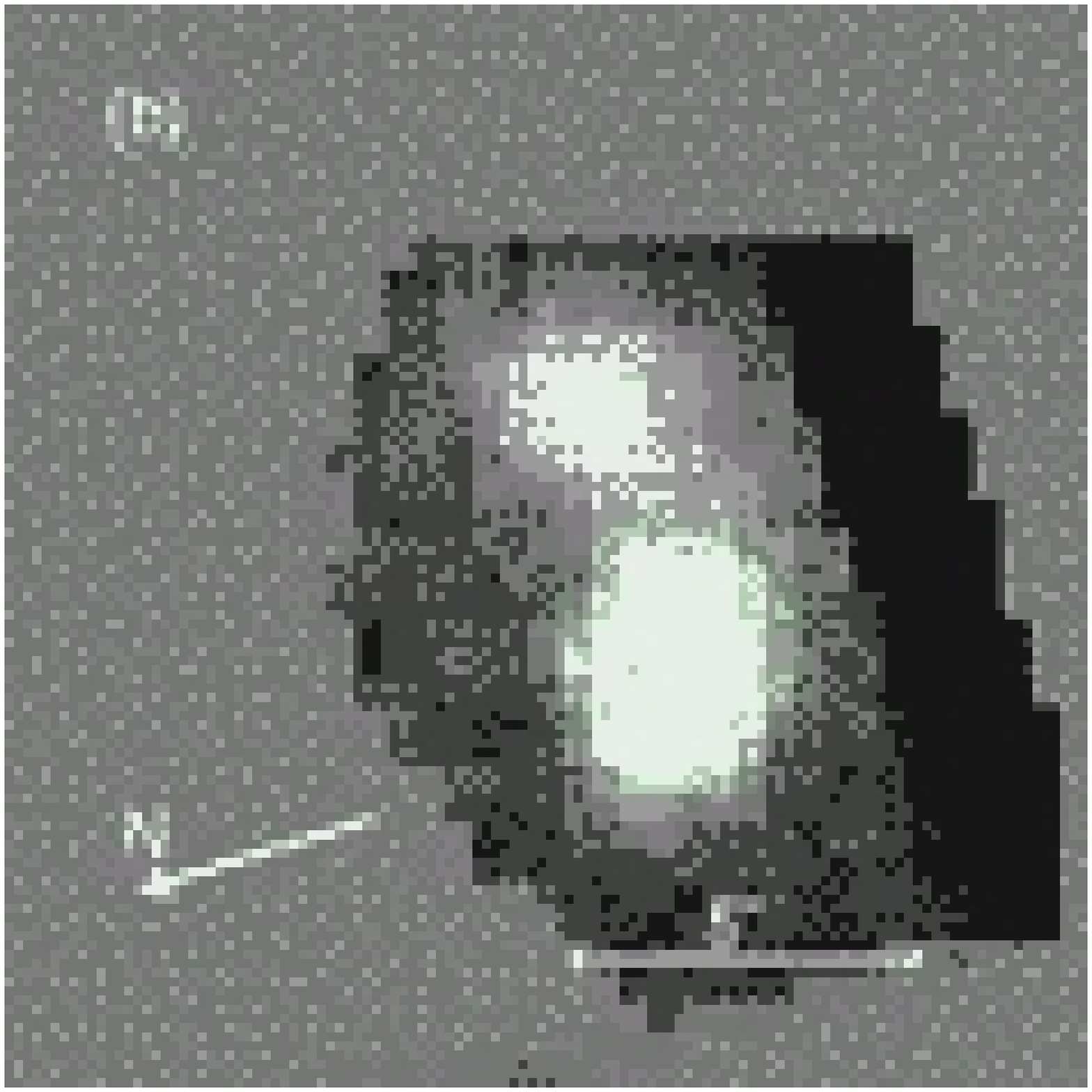}
\includegraphics[scale=0.31]{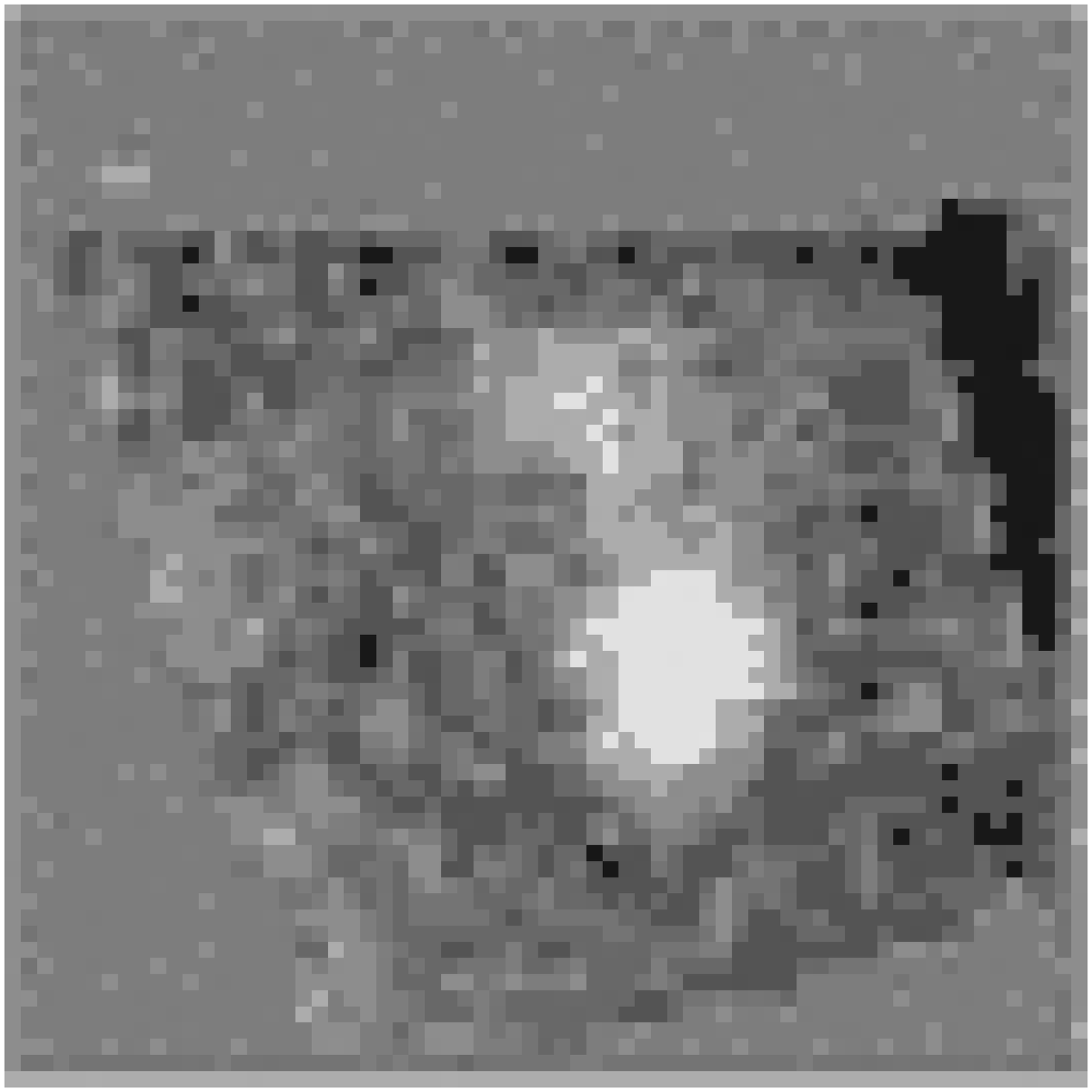}
\includegraphics[scale=0.31]{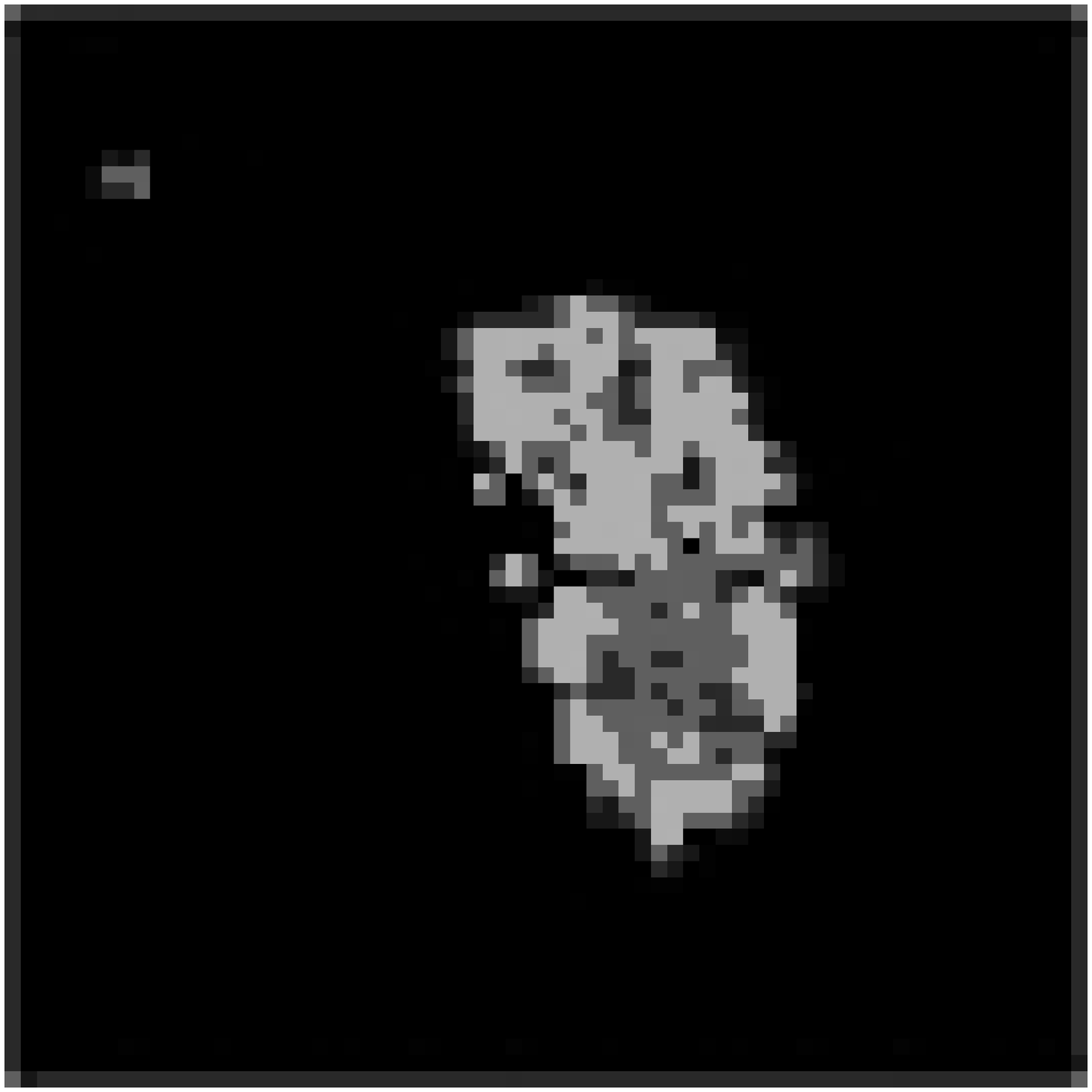}
\includegraphics[scale=0.45]{071002_k3d_radial_plot_kpc.ps_pagesj1656}
\includegraphics[scale=0.45]{071009_k3d_radial_plot_kpc.ps_pagesj1656r}
\end{center}
\caption{Figures for J1656: 
 (a) $g,r,i$-composite image taken from the SDSS website. The kyoto-3DII field-of-view is overlaid. (b) Image using the light in all the wavelengths (4200-5200\AA). Overplotted circles are with the radii of 1.5,3, and 4.5 lenslet.  The orientation of the figure is indicated with the arrow. (c) Image using the light in H$\delta$ wavelength.  The scale and orientation are common among panels (b), (c), and (d).  The effective data area is lager than that in the panel (b) because of the smaller wavelength range used. (d) H$\delta$-to-continuum ratio. 
(e)  Spectra of the central 1.5 lenslet radius region, and annuli of 1.5-3, 3-4.5 lenslets regions of the lower-right galaxy from top to the bottom.  The spectra are shifted to the rest-frame and smoothed using a 5-pixel box.  (f)  H$\delta$ (diamonds) and H$\gamma$ (triangles) EWs are plotted against the distance to the galaxy centre.
}\label{fig:J1656}
\end{figure*}

 In the panel (b) of  Figure \ref{fig:J1656}, we show the image of SDSSJ165648.64+314702.3 reconstructed using the light in all wavelength (4200-5200\AA). This galaxy, J1656, has a possible nearby companion galaxy to the west as shown in panel (a). The companion candidate is clearly detected in the panel (b) through the IFS.  

 In the panel (c), we show the image in the H$\delta$ wavelength (4086-4114\AA). In the panel (d), we show the ratio of H$\delta$ (panel c) to the continuum (panel b).  See Appendix \ref{Oct  3 10:22:16 2007} for the details in estimating the ratio when both components have large uncertainty. 
 Interestingly, in the lower-right galaxy, the strong H$\delta$ absorption (darker in the image) is concentrated around the centre of the galaxy. There is no obvious trend on H$\delta$ for the upper-left galaxy. 

 In the panel (e) of Figure \ref{fig:J1656}, we show the spectra of the central 1.5 lenslet radius region, and annuli of 1.5-3, 3-4.5 lenslet regions  of the  upper-left galaxy from top to the bottom. 
 We did not use the 4.5-6 lenslets of radius due to the smaller galaxy size.
 The spectra are shifted to the rest-frame and smoothed using a 5-pixel box.  
 Although the spectra are noisier, we recognize the same trend as J2101 in Figure \ref{fig:J2102}, i.e., both H$\delta$ and H$\gamma$ absorptions become stronger with decreasing radius.
 The trend is confirmed in the panel (f), where we show EWs of H$\delta$ and H$\gamma$ as a function of radius (from the galaxy centre). H$\delta$ EW is as large as 8\AA~at the centre, confirming our target selection of E+A galaxies. The H$\delta$ EWs stay strong as far as $\sim$3 kpc. 
 H$\gamma$ EWs are also strong at 0-3kpc annuli. Note that since the innermost spectrum is affected by a strong noise at $\sim$4300\AA, we use bluer wavelength range (4260-4280\AA) by hand in determining the continuum for H$\gamma$ EW. 
  These results suggest that the post-starburst region in E+A galaxies are extended to a few kpc of radius, and do not seem to be limited to the nuclear of the galaxy.

\begin{figure*}
\begin{center}
\includegraphics[scale=0.31]{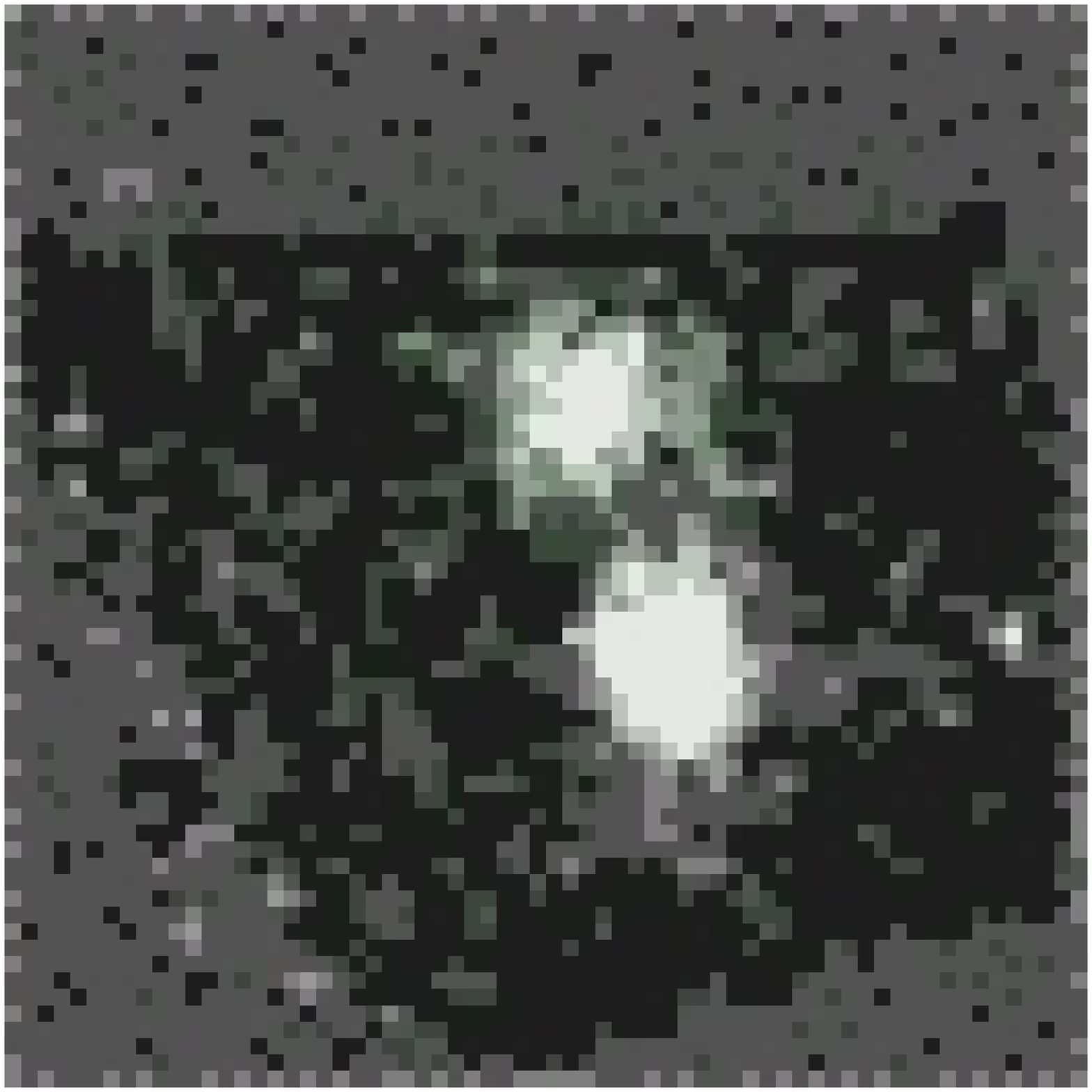}
\includegraphics[scale=0.31]{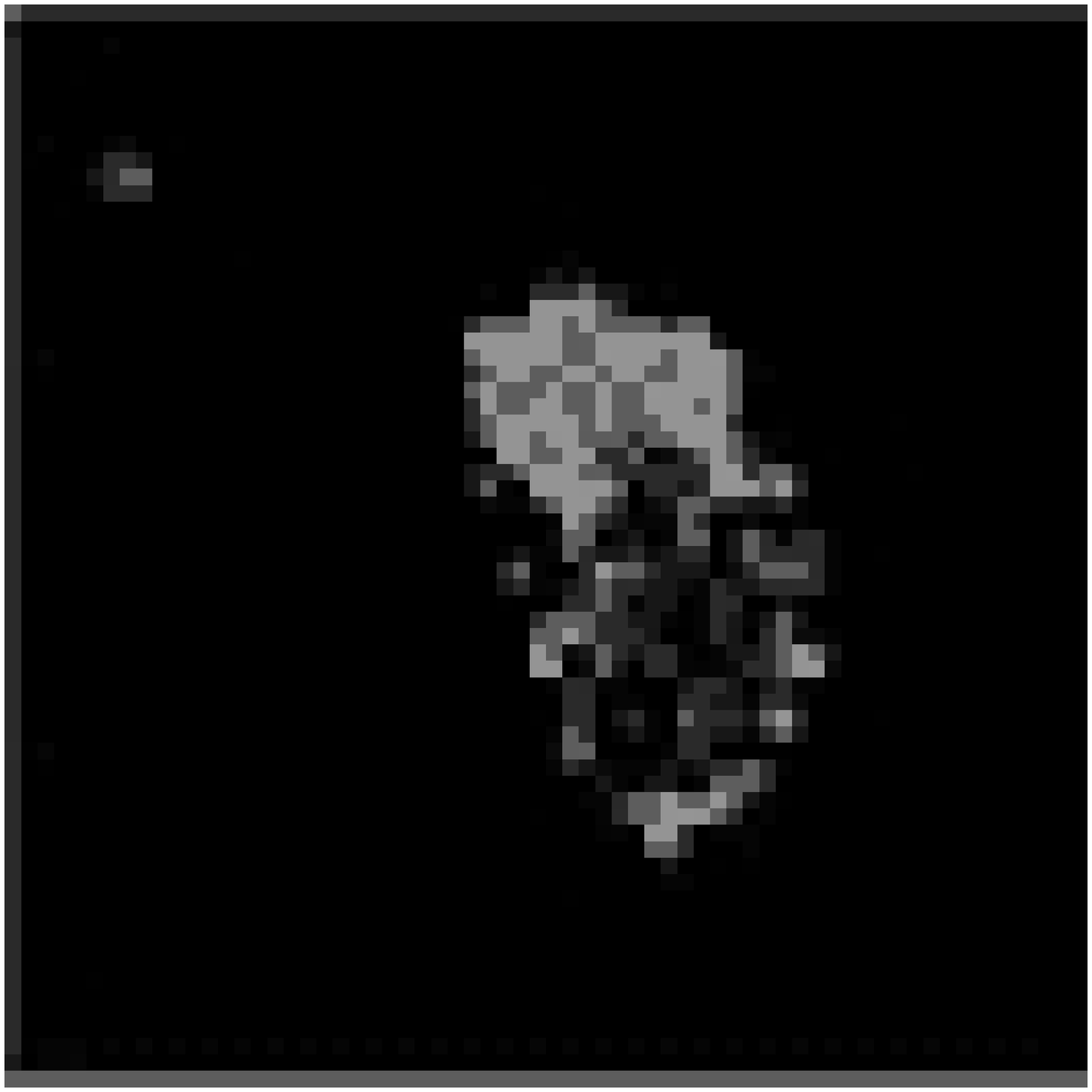}
\includegraphics[scale=0.45]{071002_k3d_radial_plot_kpc.ps_pagesj1656g}
\includegraphics[scale=0.45]{071002_k3d_radial_plot_kpc.ps_pagesj1656gr}
\end{center}
\caption{Figures for the upper-left galaxy of J1656: 
  (a)Image using the light in H$\gamma$ wavelength. Overplotted circles are with the radii of 1.5,3,4.5 and 6 lenslet. (b) H$\gamma$-to-continuum ratio. 
(c) Spectra of the central 1.5 lenslet radius region, and annuli of 1.5-3, 3-4.5 lenslet regions of the upper-left galaxy from top to the bottom.  The spectra are shifted to the rest-frame and smoothed using a 5-pixel box.  (f)  H$\delta$ (diamonds) and H$\gamma$ (triangles) EWs are plotted against the distance to the centre of the upper-left galaxy.
}\label{fig:J1656_hg}
\end{figure*}

 In the panel (a) of Figure \ref{fig:J1656_hg}, we show the image in the H$\gamma$ wavelength (4318-4354\AA). In contrast to the panel (d) of Figure \ref{fig:J1656}, the upper-left galaxy is much brighter in the H$\gamma$ wavelength. In the panel (b) of  Figure \ref{fig:J1656_hg}, we take the ratio of  H$\gamma$ to all the wavelength (See Appendix \ref{Oct  3 10:22:16 2007} for the details of the procedure). The trend is clearer, suggesting that the upper-left galaxy might have H$\gamma$ in emission.
  In the panel (c), we show spectrum of the upper-left galaxy.  As previous spectra, the spectra shown are from the central 1.5 lenslet radius region, and annuli of 1.5-3, 3-4.5 lenslet regions centred on the upper-left galaxy.
 The redshift of this galaxy turned out to be the same as the E+A galaxy in the lower-right (z=0.100). Thus, we have confirmed that this galaxy is a physically-associated companion galaxy, not a chance projection. Interestingly, this galaxy has an emission in H$\gamma$ (EW$\sim$-18.0\AA), suggesting that this galaxy is a star-forming galaxy. 
  H$\gamma$ emission of EW=-18.0\AA~is rather strong, and this galaxy may be called as a starburst galaxy.

It has been suggested that the galaxy merger might create E+A galaxies \citep[e.g.,][]{2005MNRAS.357..937G}. 
Having a close companion at the same redshift, this E+A galaxy, J1656, is a real example produced by the galaxy-galaxy merger with the upper-left companion. At the same time, J1656 presents an interesting example where the companion galaxy is not necessary an E+A galaxy, even though  it involves with the same galaxy merger. This example suggests that not only a merger but an additional condition is required to produce an E+A galaxy. 
 Only a few spectroscopic companions of E+A galaxies are known before. For example, a companion galaxy of an E+A galaxy in \citet{2006ApJ...642..152Y} was a passive galaxy. More spectroscopic follow-up of E+A companion galaxies is needed to reveal what kind of galaxy merger can produce an E+A galaxy.

%
%
%
%
%

\subsection{SDSSJ233712.76-105800.3}
\begin{figure*}
\begin{center}
\includegraphics[scale=0.33]{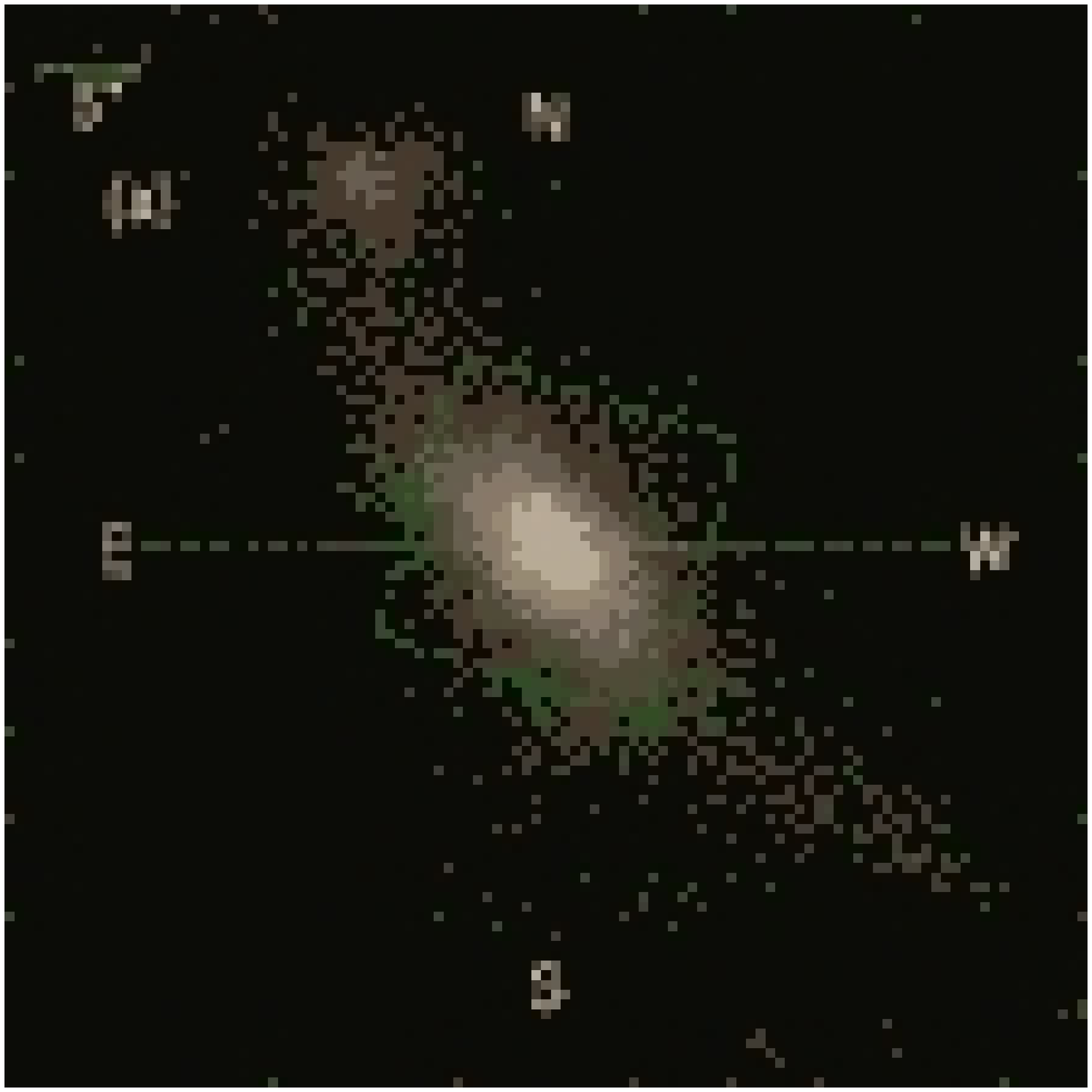}
\includegraphics[scale=0.31]{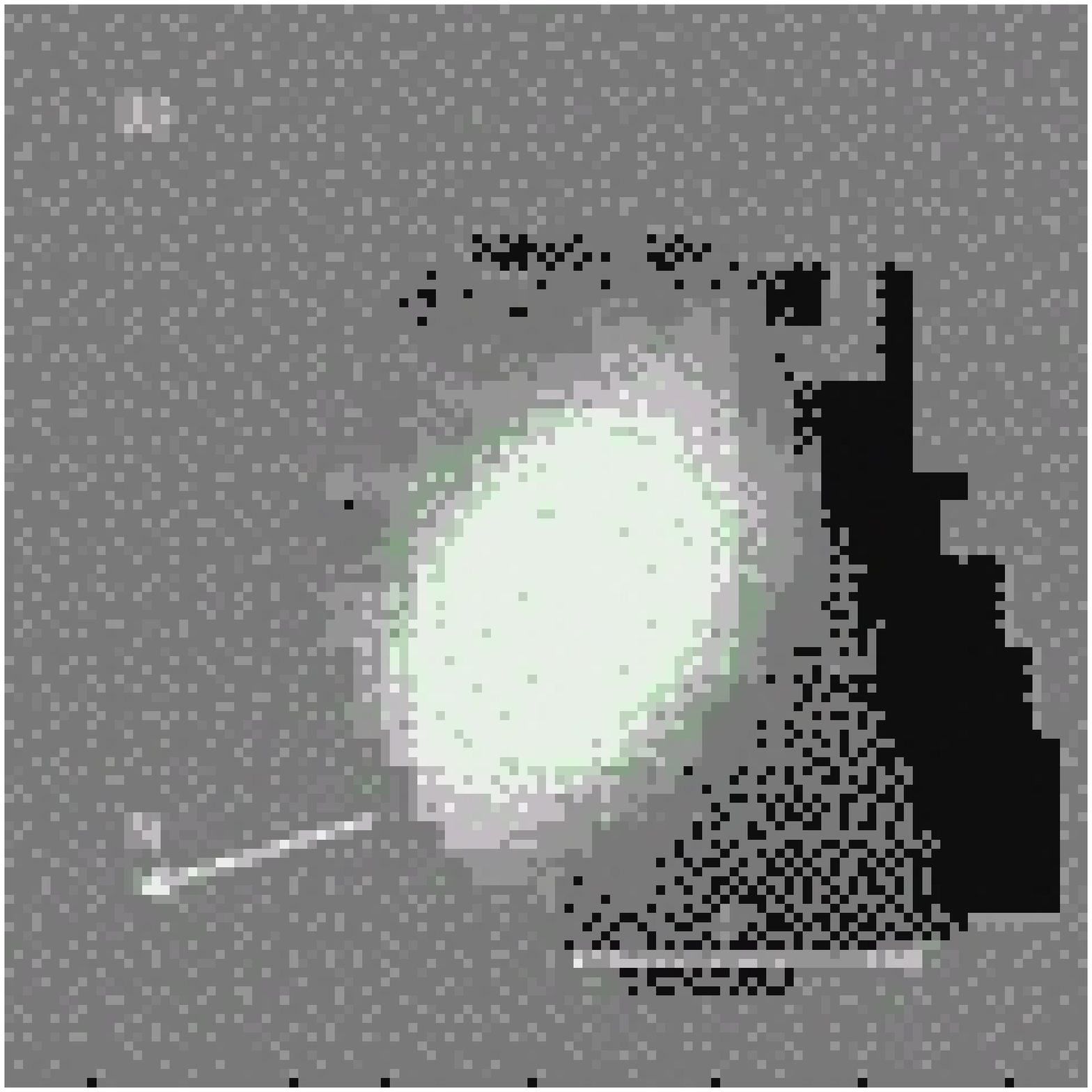}
\includegraphics[scale=0.31]{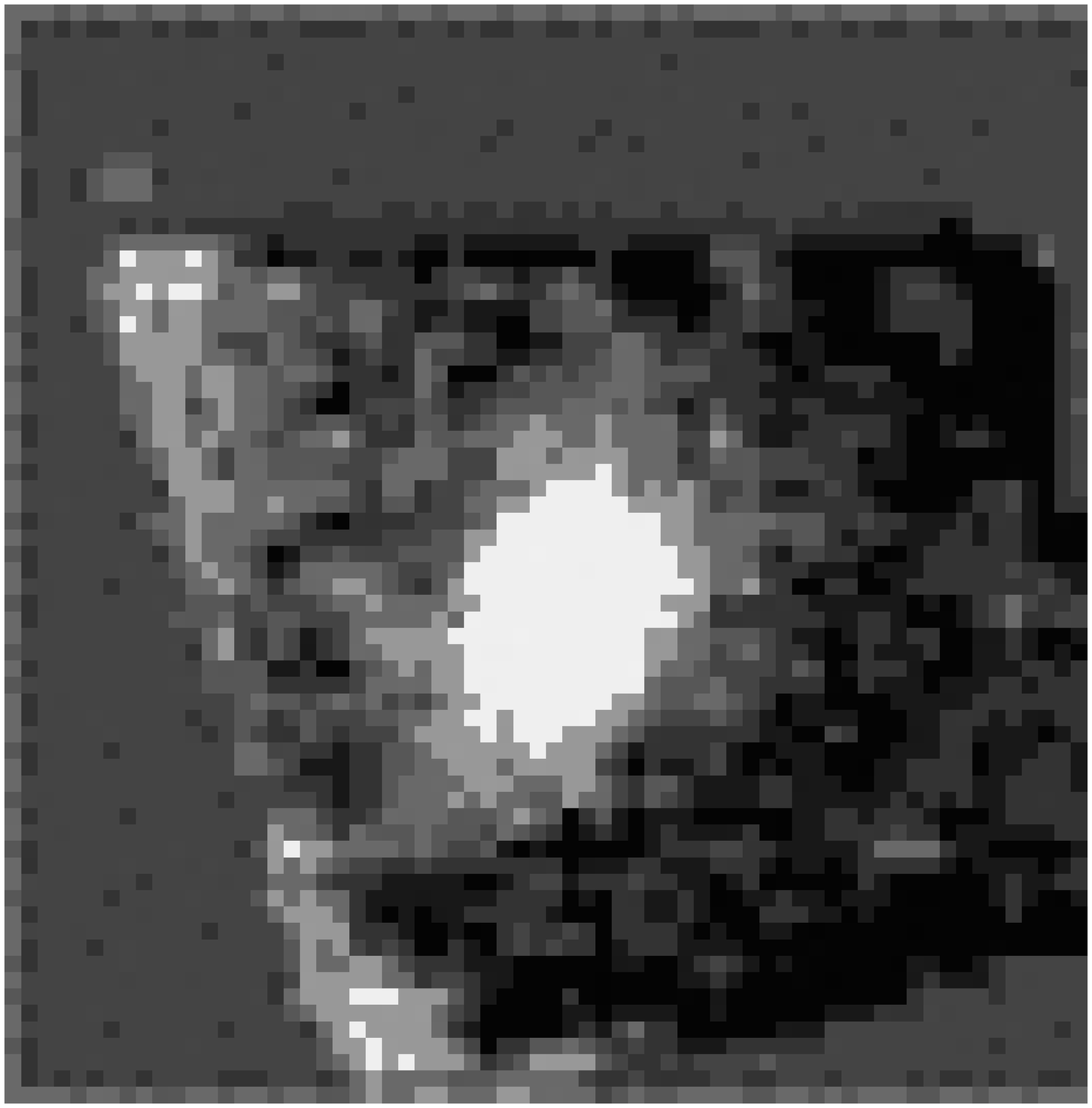}
\includegraphics[scale=0.31]{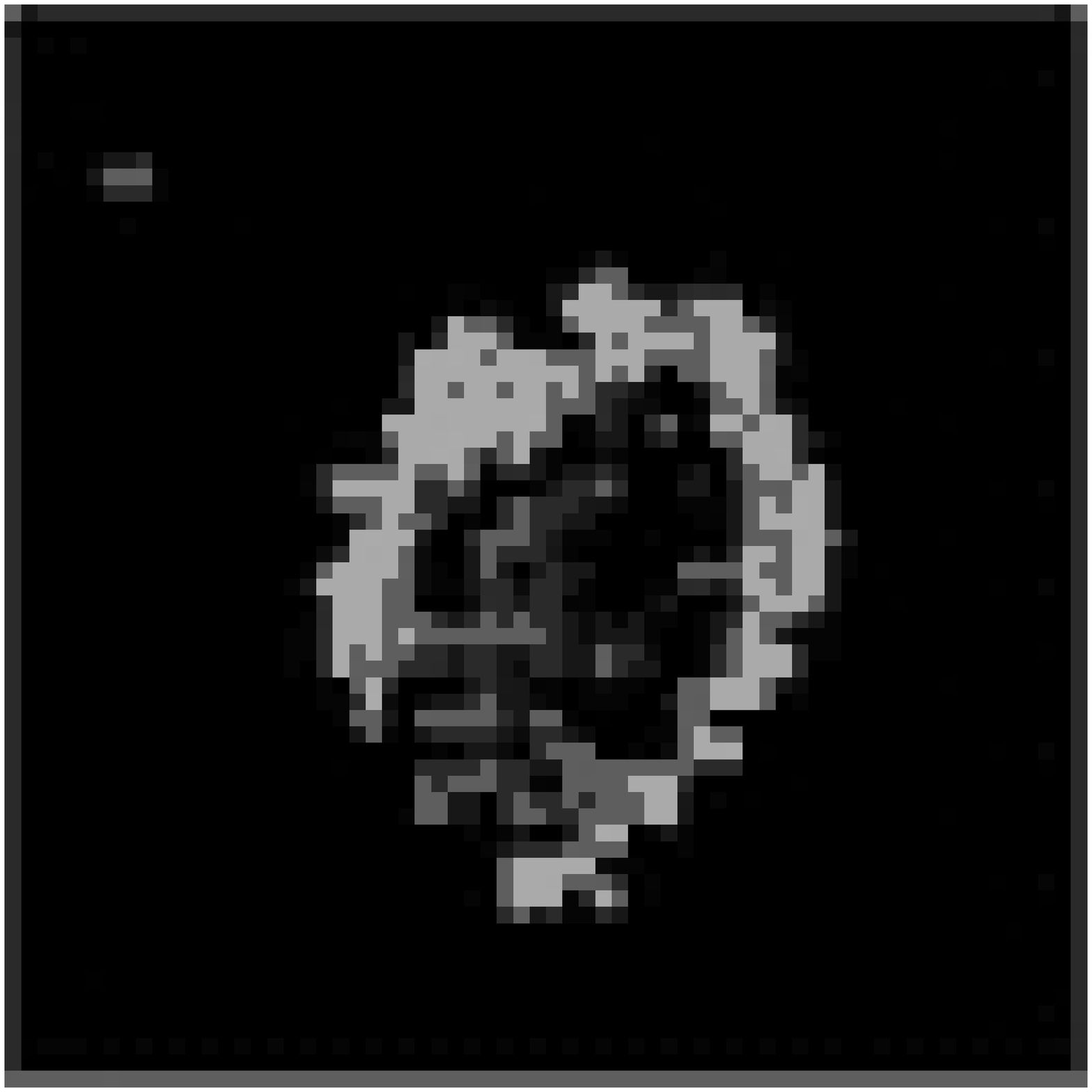}
\includegraphics[scale=0.45]{071002_k3d_radial_plot_kpc.ps_pagesj2337}
\includegraphics[scale=0.45]{071002_k3d_radial_plot_kpc.ps_pagesj2337r}
\end{center}
\caption{Figures for J2337: 
(a) $g,r,i$-composite image of J2337 taken from the SDSS website. The kyoto-3DII field-of-view is overlaid. (b) Observed image of J2337 using the light in all the wavelengths (4200-5200\AA). Overplotted circles are with the radii of 1.5,3,4.5 and 6 lenslet.  The orientation of the figure is indicated with the arrow.  (c) Image of J2337 using the light in H$\delta$ wavelength.  The scale and orientation are common among panels (b), (c), and (d).  The effective data area is lager than that in the panel (b) because of the smaller wavelength range used. (d) H$\delta$-to-continuum ratio. 
 (e)  Spectra of the central 1.5 lenslet radius region, and annuli of 1.5-3, 3-4.5, 4.5-6 lenslet regions from top to the bottom.  The spectra are shifted to the rest-frame and smoothed using a 5-pixel box.  (f) H$\delta$ (diamonds) and H$\gamma$ (triangles) EWs are plotted against the distance to the galaxy centre.
}\label{fig:J2337}
\end{figure*}

 In the panel (b) of Figure \ref{fig:J2337}, we show the images of J2337 using all the light (4200-5200\AA) through the IFS.
 Unfortunately, the tidal tail of the J2337 is out of the 15''-field-of-view of the Kyoto3DII, but the basic structures of galaxies were recovered even using the light through the IFS. 
 In the panels (c) and (d), we show the image using the H$\delta$ wavelength (4087-4119\AA) and the ratio of  H$\delta$ to the continuum as in Figs. \ref{fig:J2102} and \ref{fig:J1656}. As noticed before, H$\delta$ absorption is centrally-concentrated, but spatially-extended beyond the galaxy core. 
 Although marginally significant, we note that in the panel (d), H$\delta$ absorption may be shifted by a few lenslets toward the upper-right (or south) direction. We discuss possible interpretation of this in Section \ref{Nov  7 14:03:29 2007}.

In the panel (e) of Figure \ref{fig:J2337}, we show the spectra of the central 1.5 lenslet radius region, and annuli of 1.5-3, 3-4.5, 4.5-6 lenslets regions  of J2337 from top to the bottom. The spectra are shifted to the rest-frame and smoothed using a 5-pixel box.  The spectrum from the outermost annulus (4.5-6 lenslets of radius) is approaching to zero and not reliable. In the panel (e), it is noted that H$\delta$ and H$\gamma$ absorptions become deeper with decreasing distance to the centre of the galaxy.
 We have measured H$\delta$ and H$\gamma$ EW in these spectra in the panel (f) of Figure \ref{fig:J2337}. The H$\delta$ EWs are plotted with diamonds and the H$\gamma$ EWs are with triangles. H$\delta$ EW is the strongest at the centre with $\sim$10\AA, and remains strong as far as $\sim$2 kpc away.  H$\gamma$ EWs are not so strong as  H$\delta$, but show a similar trend.
 These results are similar to the previous two galaxies, i.e., strong H$\delta$ and H$\gamma$ absorption is observed in the core of the galaxy, and both of the lines stay strong as far as a few kpc from the centre of the galaxy.

 We would like to mention a possibly interesting different radial trend of  H$\delta$ and H$\gamma$ in the panel (f);  H$\delta$ EW is the strongest at the innermost bin, and declines immediately at the 2nd bin to stay flat at the 3rd bin. On the other hand,  H$\gamma$ EW in the 2nd bin is as large as in the 1st bin, and suddenly declines at the 3rd bin. These behaver can also be seen for J2102 in the panel (f) of Fig. \ref{fig:J2102}. We will discuss possible interpretation in Appendix \ref{Nov 15 12:43:11 2007}.


%
%

\section{Discussion}
\label{Nov  7 14:03:29 2007}

 One of the main purposes of this paper is to shed light on the origin of the E+A galaxies based on the internal properties of E+A galaxies obtained with the Kyoto3DII IFS. Previous work on E+A galaxies was noticeably focused on the global/external properties of the E+A galaxies such as integrated spectra \citep{1999ApJ...518..576P} and environment\citep{2005MNRAS.357..937G}, and thus, it is important to obtain independent evidence on the origin based on the internal structure.

 We have observed three E+A galaxies with the Kyoto3DII IFS, with the main results shared among these three galaxies; Balmer absorption, especially H$\delta$ absorption, was strongest at the centre/core but significantly-extended in a few kpc scale in all the three galaxies, suggesting that the post-starburst phenomena are centred on the galaxy core, but not limited to the core. These results have an important physical implication on the origin of E+A galaxies.

\citet{2005MNRAS.359.1421P} has run numerical simulations of the galaxy merger model and the truncation model (e.g., the tidal stripping or the ram-pressure stripping model) to predict radial profiles of the H$\delta$ EW in the frame work of \citet{1998ApJ...497..108B,2001ApJ...547L..17B,2002ApJ...577..651B}. 

\begin{itemize}
 \item 
 In their galaxy-galaxy merger simulation, a centralized burst of star formation is produced, when the starburst ends the galaxy is left with a central population of young stars and hence a radial distribution of H$\delta$ EW which is highest in the centre and decreases rapidly with galactocentric radius. After $\sim$1.5 gigayears, the radial H$\delta$ profile evolves to be flat and uniformly low across the entire extent of the galaxy. 
 This result is a natural consequence of recent major-merger computer simulations, in which during galaxy-galaxy collisions, the gas readily loses angular momentum due to dynamical friction, decouples from the stars, and inflows rapidly toward the merger nuclei \citep{1992ApJ...393..484B,1992ARA&A..30..705B,1996ApJ...471..115B,1994ApJ...425L..13M,1996ApJ...464..641M}, creating a central starburst, which could evolve into E+A phase if the truncation of the starburst is rapid enough \citep{2004A&A...427..125G}.

 \item 
 Their truncation model assumes a situation like a ram-pressure stripping model or a tidal stripping model, in which the star-formation is simultaneously and uniformly truncated throughout the entire disc. Immediately after the truncation of star formation, the galaxy has a flat, uniformly high $H\delta$ EW profile. As the system evolves, the contribution from the old stellar-population becomes most prominent at the centre of the galaxy; this results in the H$\delta$ EWs decreasing most rapidly in the central region of the galaxy. 
\end{itemize}

 A stark contrast between these two models makes the H$\delta$ profile a powerful tool to investigate the physical origin of E+A galaxies. In the panel (d) of Figures \ref{fig:J2102},\ref{fig:J1656} and \ref{fig:J2337}, we observed that a strong H$\delta$ absorption is centred around the galaxy core in all the three cases. Although the relative strength of the H$\delta$ absorption depends on the strength of the starburst and the time since then, the observed trend is qualitatively consistent with the prediction from the galaxy-galaxy merger simulation. This results are unique in that it is obtained using only the internal structures of E+A galaxies.

 Previous attempts also support our results: 
  \citet{1996AJ....111...78C} obtained long-slit spectra of E+A galaxies in the Coma cluster and showed that starburst signatures are prominent in the central core and are spatially extended. 
  Similarly, \citet{2001ApJ...557..150N} found that  the young stellar populations are more centrally concentrated than the older populations, but they are not confined to the galaxy core (radius $<$1 kpc).  
 Recently, \citet{2006AJ....131.2050Y} performed spatially-resolved spectroscopy of three E+A galaxies using the APO3.5m telescope. They found that H$\delta$ EWs were the largest at the galaxy centre although the strong H$\delta$ was significantly extended toward outside of the galaxies ($>4$kpc). 
  Later,  \citet{2006ApJ...642..152Y} observed a nearby E+A galaxy with a red companion galaxy at the same redshift (z=0.033) with the FOCAS long-slit spectrograph on the Subaru telescope. The spatially-resolved spectra showed that the H$\delta$ EWs were also strongest at the centre, but extended over $\sim$5kpc. 
 \citet{2005ApJ...622..260S} observed a H$\delta$-strong galaxy with a weak [OII] emission  from the \citet{2003PASJ...55..771G} catalogue, and found that A stars are widely distributed across the system and are not centrally concentrated. Note, however, that this galaxy had a weak emission in [OII] (4.1\AA), and thus, cannot be called as an E+A galaxy in our definition. 
 \citet{2005MNRAS.359.1557Y} found that a significant number of E+A galaxies exhibit positive slope of radial colour gradient (bluer gradient toward the centre), being consistent with the hypothesis that E+A galaxies are caused by merger/interaction, having undergone a centralized violent starburst. 

 All these work seems to be still affected by random noises possibly coming from sample selections, and possibly from the errors in the measurement.
 However, majority of work including this work seem to be conversing to a certain direction, i.e., the post-starburst phenomena is centrally-concentrated in most cases, and significantly extended to a few kpc, being consistent with a theoretical prediction for merger/interaction remnant. 
 These results are unique in that without using information on the global/external properties of E+A galaxies, but yet reached a similar conclusion on the origin of E+A galaxies. For example, \citet{2005MNRAS.357..937G} found an excess in the number of companion galaxies of E+A galaxies, concluding that it is an evidence of the merger-interaction. Many authors reported disturbed morphologies of E+A galaxies point to the merger/interaction origin \citep{1991ApJ...381L...9O,2004MNRAS.355..713B,2006astro.ph.12053L}.
 
 E+A galaxies have often been thought to be altered by the ram-pressure stripping in galaxy cluster environment, as represented by the Butcher Oemler effect \citep[e.g.,][]{1978ApJ...226..559B,2003PASJ...55..739G,2003PhDT.........2G}. However, our results suggest that the ram-pressure stripping may not be the cause. Passive spiral galaxies \citep{1998ApJ...497..188C,1999ApJ...518..576P,2003PASJ...55..757G,2004MNRAS.352..815Y} may be an alternative victim of the ram-pressure stripping in galaxy cluster environments.

 Regarding the work that found a flat, not centrally-concentrated H$\delta$ radial profile \citep{2005MNRAS.359.1421P,2005ApJ...622..260S}, these imply that there might be multiple physical mechanisms to create an E+A galaxy (e.g., the ram-pressure stripping). Especially, E+A galaxies in \citet{2005MNRAS.359.1421P} are in high-redshift galaxy cluster, and thus, there is a significant environmental difference from our sample of field E+A galaxies.  
Note that, however, these samples often lacked information on [OII] or H$\alpha$ emission line, stressing importance in careful sample selection requiring no emission in both [OII] and H$\alpha$.
 Another possible reason is the luminosity difference. In the last column of the Table \ref{tab:targets}, we list absolute magnitudes of our target galaxies in $r$-band. The k-correction is applied using \citet[][v\_3.2]{2003ApJ...594..186B}. The absolute magnitudes range from $-$21.7 to $-$23.0, which are brighter than the $M^*_r$(=-21.2 in our cosmology) of local galaxies \citep{2003ApJ...592..819B}. The reason why our target galaxies are luminous is partly because of our selection of apparently brighter galaxies and partly because of the flux-limited nature of the SDSS survey. However, it is also possible that nearby E+A galaxies may be intrinsically luminous. It is important to investigate the luminosity/mass distribution of local E+A galaxies, and such a study is under progress (Inami et al. in preparation).
 The galaxy in \citet{2005ApJ...622..260S} is taken from our sample\citep{2003PASJ...55..771G,goto_DR6}, and has an absolute magnitude of $M_r$=-22.32. The galaxies in  \citet{2005MNRAS.359.1421P} have 18.4$<R<$20.3 at $z=0.32$. Therefore, the expected absolute magnitudes range is  -21.4$<M_r<$-20.5 (the colour conversion in \citet{1995PASP..107..945F} was used), and thus they are significantly less luminous galaxies, which may be more vulnerable to the gas stripping in the cluster environment due to their smaller gravitational potential. An IFS observation of less luminous nearby E+A galaxies would shed more light on the subject.
 
 In panels (d) of Figs.\ref{fig:J2102},\ref{fig:J1656} and \ref{fig:J2337}, we showed the ratio of H$\delta$-to-continuum in the 2D images. Interestingly, panels (d) of Figs. \ref{fig:J2102} and \ref{fig:J2337} show a possible shift of H$\delta$ absorption toward south-west (although it is centrally-concentrated when radially averaged);
 In the panels (d) of Fig.  \ref{fig:J2337}, the darkest region inside a ring-like structure at the centre of the galaxy is slightly shifted toward the  south-west direction.
 No shift is observed in the panel (d) of Figure \ref{fig:J1656} for J1656. Curiously, the directions of the shifts observed for J2102 and J2337 coincide with those of the tidal tails seen in the panels (a). In the panel (a) of Fig. \ref{fig:J2102}, there is a possible tidal tail extended toward south-west direction. In case of  the panel (a) of Fig. \ref{fig:J2337}, there exists an obvious tidal tail from the north-east to south-west. The coincidence of these post-starburst regions with the direction of the tidal tails suggests that the dynamical merger/interaction might have played an important role in forming the post-starburst regions in these galaxies. Although details of the coincidence need to be verified with the numerical simulations,  this demonstrates the power of the Kyoto-3DII in investigating the 2D structure of the post-starburst regions.

 We have to be careful on possible selection effect; our E+A selection is based on the SDSS fiber spectrograph which has a diameter of three arcsec, i.e., by default, our E+A galaxies have strong H$\delta$ absorption at the centre of the galaxy. Therefore, we do not find a galaxy with a poststarburst ring, or with a post-starburst region only at the edge of a spiral arm in our sample. This bias is especially strong for nearby galaxies at $z<0.05$. Fortunately in this work, due to the wavelength coverage of the instrument, the observed galaxies have a larger redshift of 0.078,0.093 and 0.100, and thus, are relatively immune from this aperture bias \citep[see Fig.5 of][]{2003MNRAS.346..601G,goto_DR6}. In addition, immediately after the truncation in the stripping model, the simulation predicts a flat H$\delta$ profile, which can be very well found in our sample. No finding of such E+A galaxies in our sample still carry valuable information on the origin of E+A galaxies.

\section{Conclusions}
 We have observed three nearby E+A galaxies with the Kyoto3DII IFS. 
 The spatially-resolved spectra show that the strong Balmer absorption characteristic to E+A galaxies is concentrated around the galaxy centre, but is spatially-extended to a few kpc scale. 
 These results support a scenario where the infalling gas caused by galaxy-galaxy merging may create the post-starburst phase in the centre of the galaxy.
 We have found that the extensions of the post-starburst regions coincide with the direction of the tidal tails for J2102 and J2337, possibly supporting the merger/interaction scenario.
 We have confirmed that a nearby object to J1656 was at the same redshift, and thus is physically-associated. However interestingly, this nearby companion is a star-forming galaxy, and not an E+A galaxy. This implies that the galaxy-galaxy merging may create E+A galaxies at a certain condition, but not all  merging galaxies will evolve into an E+A galaxy. 
 We have found that the H$\delta$ and H$\gamma$ absorption lines show different radial trends, which were difficult to be interpreted with the age/metallicity mixture. 


\section*{Acknowledgments}

We thank the anonymous referee for many insightful comments, which significantly improved the paper.
We are grateful to Yoshiko Okita for valuable help in preparing the observation.
We thank Chisato Yamauchi for useful discussions. 


 Use of the UH 2.2-m telescope for the observations is supported by NAOJ.
 The research was financially supported by the Sasakawa Scientific Research Grant from The Japan Science Society.
 This research was partially supported by the Japan Society for the Promotion of Science through Grant-in-Aid for Scientific Research 18840047.

    Funding for the creation and distribution of the SDSS Archive has been provided by the Alfred P. Sloan Foundation, the Participating Institutions, the National Aeronautics and Space Administration, the National Science Foundation, the U.S. Department of Energy, the Japanese Monbukagakusho, and the Max Planck Society. The SDSS Web site is http://www.sdss.org/.

    The SDSS is managed by the Astrophysical Research Consortium (ARC) for the Participating Institutions. The Participating Institutions are The University of Chicago, Fermilab, the Institute for Advanced Study, the Japan Participation Group, The Johns Hopkins University, Los Alamos National Laboratory, the Max-Planck-Institute for Astronomy (MPIA), the Max-Planck-Institute for Astrophysics (MPA), New Mexico State University, University of Pittsburgh, Princeton University, the United States Naval Observatory, and the University of Washington.


\def\Hg{H$\gamma$}
\def\Hd{H$\delta$}

\appendix

\section{division of two low signal-to-noise ratio images}
\label{Oct  3 10:22:16 2007}

\begin{figure}
\includegraphics[scale=0.3,angle=-90]{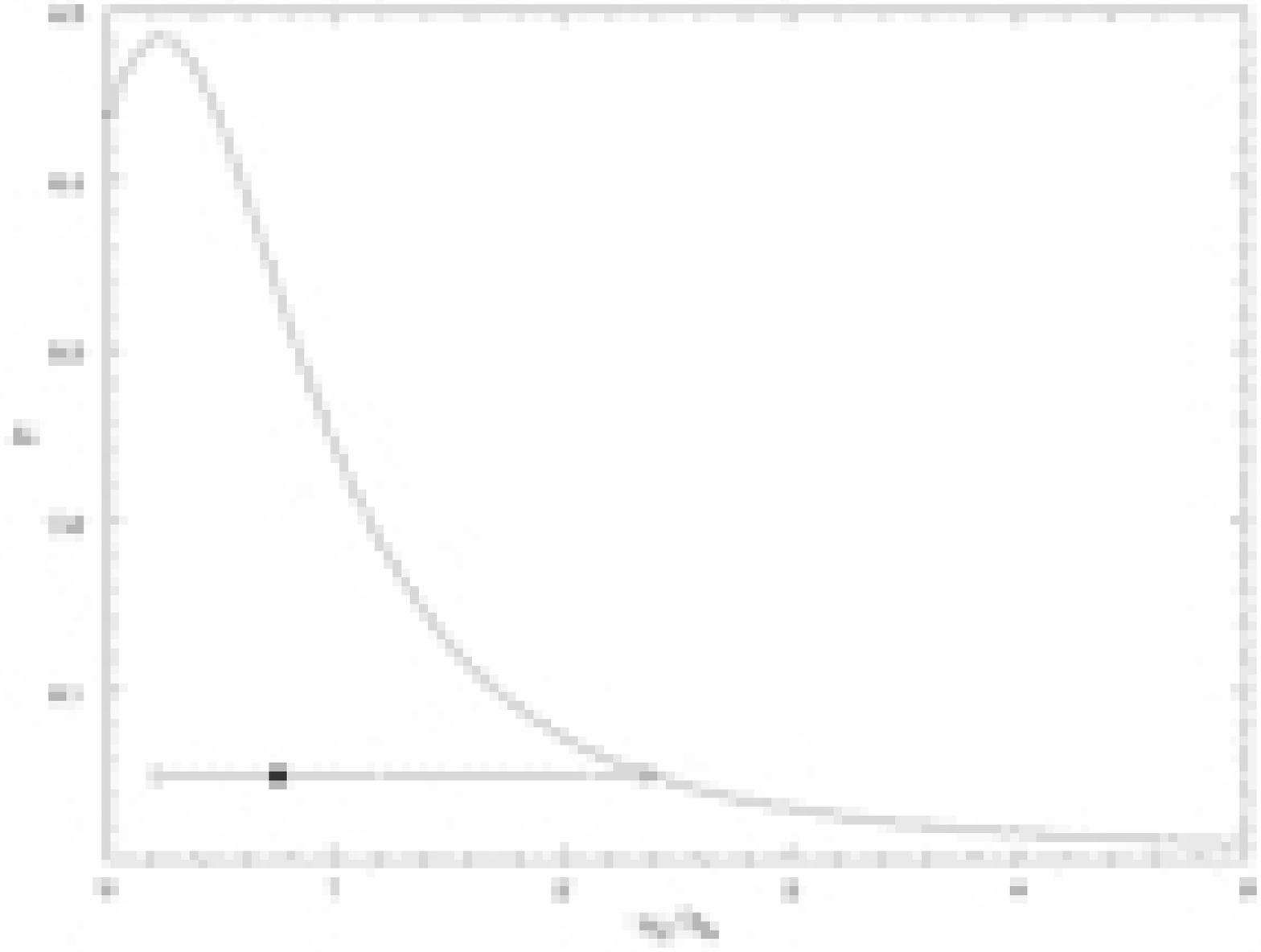}
\caption{An example of estimation of distribution
of $a_0/b_0$, where  a=0.5, b=1.0, and $\sigma_a$=$\sigma_b$=1.
The dot represents the median and the error bar represents
the upper and the lower 15\% percentile.
}
\end{figure}

In the panel (c) of Figures \ref{fig:J2102}-\ref{fig:J2337}, we have shown the ratio of the H$\delta$ iamge to the continuum image. However, a straight division of two images does not return a correct ratio since both images have large uncertainty. In this section, we explain the details of the method we used to derive the ratio.
 
We estimate \Hd\ -to-continuum ratio in 
low S/N region as follows. 
We call \Hd\, (or \Hg\ ) image as A, and continuum image 
as B, and estimate A/B.
The observed value of image A and B
at a certain pixel ($a$ and $b$, hereafter)
follow some distribution whose typical values are
$a_0$ and $b_0$.
The problem is how to estimate $a_0/b_0$ from $a$, $b$ 
and other information.

Formalizing the problem, we assume that the $a$ and $b$
follow the normal distribution with variance 
$\sigma_a^2$ and $\sigma_b^2$.
The variance of each distribution is estimated, 
assuming that the primary source of the variation are 
 the sky noise and readout noise.

The prior distribution of $a_0, b_0$ are estimated as
\begin{eqnarray}
P(a_0) d a_0 &=&  \frac{1}{\sqrt{2\pi}\sigma_a} 
exp\left( -\frac{(a_0-a)^2}{2 \sigma_a^2}\right) d a_0,\\
P(b_0) d b_0 &=& \frac{1}{\sqrt{2\pi}\sigma_b} 
exp\left( -\frac{(b_0-b)^2}{2 \sigma_b^2}\right) d b_0
\end{eqnarray}
The prior distribution of ratio, 
$r=a_0/b_0$, is then calculated as,
\[
P(r) dr =\frac{1}{2\pi\sqrt{\sigma_a \sigma_b}}
\left(\int_0^{\infty} x
exp \left(\frac{(x-b)^2}{2\sigma_b^2}\right) 
exp \left(\frac{(r x - a)^2}{2\sigma_a^2}\right) dx\right) dr
\]
The distribution is not the normal distribution.
We therefore adopted median as the typical value of
the distribution and take the upper and the 
lower 15\% percentile as the error of the estimation.

An example of the distribution is shown in Figure A1.
When $a$=0.5, $b$=1.0, and $\sigma_a$=$\sigma_b$=1,
the estimated ratio is $0.75_{-0.53}^{+1.62}$.
It should be noted that simple division, $a/b=0.5$ is
not the best estimation for $a_0/b_0$.


\section{A systematic test: an Equivalent width profile of a standard star}
\label{Oct  3 10:21:43 2007}

\begin{figure}
\begin{center}
\includegraphics[scale=0.6]{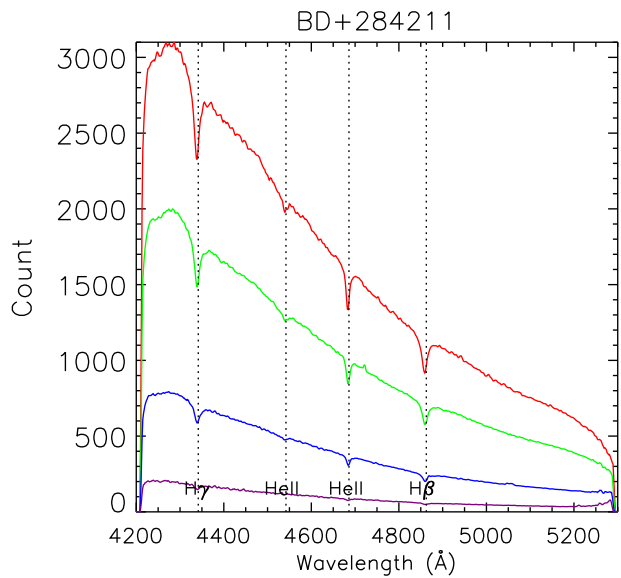}
\end{center}
\caption{ Spectra of a standard star, BD+284211, in the central 1.5 lenslet radius region, and in the annuli of 1.5-3, 3-4.5, 4.5-6 lenslets regions from top to the bottom. The spectra are smoothed using a 5-pixel box. 
}\label{fig:bd284211_spectra}
\end{figure}

\begin{figure}
\begin{center}
\includegraphics[scale=0.6]{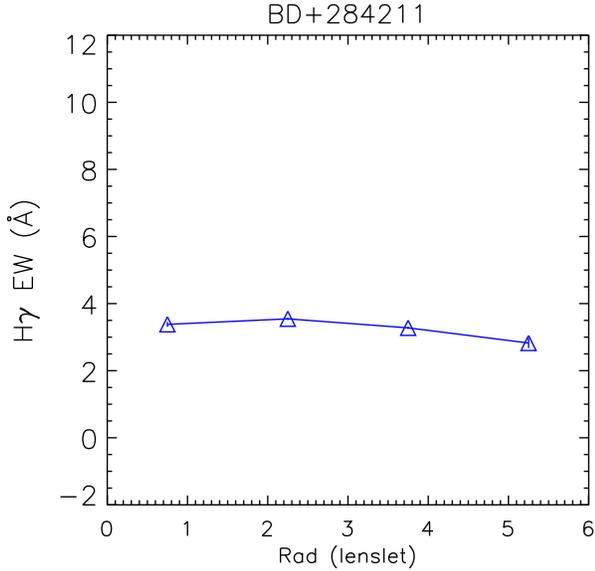}
\end{center}
\caption{
 H$\gamma$ EWs (triangles) are plotted against the distance (in the unit of lenslet) from the centre of the star.
 }\label{fig:bd284211_ew}
\end{figure}

 In Section \ref{Sep 28 14:10:45 2007}, we showed that H$\delta$/H$\gamma$ EW were larger at the centre of E+A galaxies, but extended to a few kpc scale. However, it is important to check if this result is not caused by any systematic effect. In this section, we perform the same analysis as Section  \ref{Sep 28 14:10:45 2007} on a standard star,  BD+284211, which we observed right before the galaxy J2102. The observed position of the standard star is very close to that of J2101 (within a few lenslets), and therefore, the star allows us to check any position-dependent systematic effect on the lenslet if any. 

 In Figure \ref{fig:bd284211_spectra}, we show spectra of  BD+284211 in the central 1.5 lenslet radius region, and in the annuli of 1.5-3, 3-4.5, 4.5-6 lenslet regions from top to the bottom in exactly the same way as in Figs.\ref{fig:J2102},\ref{fig:J1656} and \ref{fig:J2337}.  The spectra are smoothed using a 5-pixel box. 
 H$\beta$, H$\gamma$, and HeII (4542, 4686\AA) lines are marked.
 
 In Figure \ref{fig:bd284211_ew}, we show the H$\gamma$ EW of each spectrum as a function of radius (from the centre of the star). H$\delta$ line of the star is not covered with our grism/filter. The figure shows that the H$\gamma$ EW does not depend on the position of the lenslet, and stays at $\sim$3.3\AA.~ This test demonstrates that the radial trends we observed in Figs. \ref{fig:J2102},\ref{fig:J1656} and \ref{fig:J2337} are not caused by any position-dependent systematic effect of the lenslet.

\section{Possible different radial change of H$\delta$ and H$\gamma$ equivalent widths}
\label{Nov 15 12:43:11 2007}

\begin{figure}
\begin{center}
\includegraphics[scale=0.4]{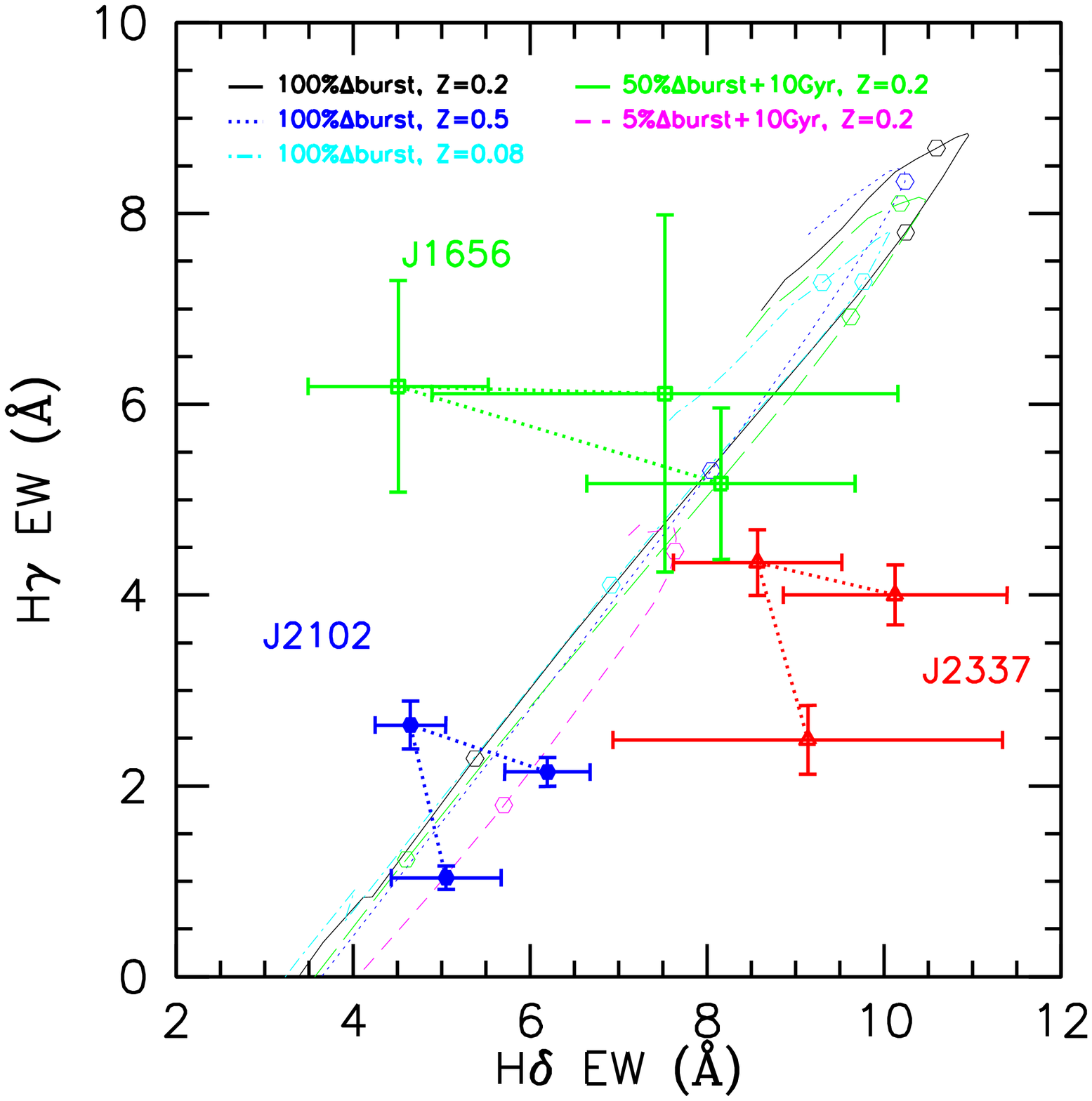}
\end{center}
\caption{H$\gamma$ EW is plotted against H$\delta$ EW for inner three data points of J2102 (filled circles), J1656 (squares) and J2337 (triangles). The solid (black) and  dotted (blue) and dot-dashed (cyan) lines are SED models \citep{2003MNRAS.344.1000B} with an instantaneous burst with Z=0.2,0.5 and 0.08, respectively.  The long-dashed (green) and dashed (pink) lines show mixture models where 5 or 50\% mass of instantaneous burst population is added on the 10Gyr-old stellar population. Open circles on the models show age of 250, 500 and 1000 Myrs, where the lines connect ages from 100 Myr to $>2$Gyrs. 
}\label{fig:hdhg}
\end{figure}

 We would like to mention an interesting radial change of H$\delta$/H$\gamma$ EWs; In the panel (d)s of Figs. \ref{fig:J2102} and \ref{fig:J2337}, H$\delta$ EW is highest at the 1st (innermost) bin, and then declines immediately at the 2nd bin to stay flat towards the 3rd bin (2-3 kpc). On the other hand,  H$\gamma$ line keeps large EWs at the 1st and 2nd bins, and then suddenly declines at the 3rd bin.  Although we do not include J1656 in this discussion due to the larger errors, the feature is common to two galaxies (J2102 and J2337), and thus, may originate from the nature of E+A galaxies. 
   H$\delta$ and H$\gamma$ EWs have been known to decrease at a similar rate with increasing age \citep[e.g.,][]{1999ApJS..125..489G}. Therefore, it is interesting that the observational data show the different decline rates for H$\delta$ and H$\gamma$ EWs as a function of galaxy radius. 

To interpret this trend, we plot H$\delta$ vs H$\gamma$ EWs for J2102, J1656, and J2337 with filled dots, squares, and triangles, respectively in Figure\ref{fig:hdhg}. Because the errors are larger than radial trends, we do not include J1656 in the following discussion. Overplotted lines are \citet{2003MNRAS.344.1000B} models with the age of 100-2000 Myrs. The solid (black), dotted (blue) and dot-dashed (cyan) lines are for instantaneous burst models with metallicity of Z=0.2(solar),0.5 and 0.08, respectively. The Salpeter initial mass function was assumed. These three models are almost on the same line (roughly speaking, H$\gamma$=1.2$\times$H$\delta-$4.3) and show that metallicity does not affect the  H$\delta$/H$\gamma$ relation significantly. The long-dashed (green) and dashed (pink) lines show hybrid models where 5 or 50\% mass of instantaneous burst population is added on the 10Gyr-old stellar population. Open circles on the models show ages of 250, 500 and 1000 Myr.  The 5\% burst model show a little deviation from the other models due to the presence of the old stellar population. However, the deviation is only by 1\AA, which is not significant enough to explain the observed change of the H$\delta$/H$\gamma$ relation (by $>2\AA$). 

 In summary, the observed trend in the H$\delta$/H$\gamma$ relation is difficult to be reproduced in the current framework of the SED models neither by the metallicity change or by the mixture of stellar populations of different age. The only large change in H$\delta$/H$\gamma$ can be brought by the age. But the change is limited on a narrow locus along the models (H$\gamma$=1.2$\times$H$\delta-$4.3), which do not agree with the observed data. To fully understand the observed radial change, both more sophisticated SED models and more IFS data with better signal-to-noise ratio are required.




\label{lastpage}

\end{document}